\documentclass[prd,aps,floats,twocolumn,nofootinbib]{revtex4}
\usepackage{commath}
\pdfoutput=1
\usepackage[usenames, dvipsnames]{color}
\usepackage{graphicx, array}
\usepackage{multirow}
\usepackage{graphics}
\usepackage{graphics,epsfig}
\usepackage{amssymb}
\usepackage{amsmath}
\usepackage{ulem}
\usepackage{multirow}
\usepackage{subfigure}
\usepackage{bm}
\usepackage{txfonts}
\usepackage{cancel}
\usepackage{url}
\usepackage{hyperref}
\usepackage{ulem}
\newcolumntype{C}[1]{>{\centering\arraybackslash}m{#1}}

\def\be{\begin{equation}}
\def\ee{\end{equation}}
\def\bi{\begin{itemize}}
\def\ei{\end{itemize}}
\def\ben{\begin{enumerate}}
\def\een{\end{enumerate}}
\def\bt{\begin{tabular}}
\def\et{\end{tabular}}
\def\bc{\begin{center}}
\def\ec{\end{center}}

\def\bea{\begin{eqnarray}}
\def\eea{\end{eqnarray}}

\def\ba{\begin{eqnarray}}
\def\ea{\end{eqnarray}}

\let\oldhat\hat
\renewcommand{\vec}[1]{\boldsymbol{\mathbf{#1}}}
\renewcommand{\hat}[1]{\oldhat{\boldsymbol{\mathbf{#1}}}}

\def\|{\ |\ }
\def\p{\parallel}

\graphicspath{ {./images/}}

\begin{document}

\input{epsf}

\title{Challenges in Constraining Gravity with Cosmic Voids}
\author{Christopher Wilson and Rachel Bean}

\affiliation{Department of Physics, Cornell University, Ithaca, New York 14853, USA.}
\affiliation{Department of Astronomy, Cornell University, Ithaca, New York 14853, USA.}

\begin{abstract}

We compare void size and clustering statistics for nDGP and $f(R)$ gravity models and GR using N-body simulations. We show how it is critical to consider the statistics derived from mock galaxy catalogs rather than the dark matter halos alone. Marked differences between the void size functions for GR and $f(R)$ models which present when voids are identified using dark matter halos are removed when voids are identified, more realistically, from mock galaxy tracers of the halos. The void radial velocities and velocity dispersions in the $f(R)$ and nDGP models are enhanced relative to GR in both halos and mock galaxy identified voids. Despite this, we find that the redshift space void quadrupole moments derived from the mock galaxy tracers are strikingly similar across the three gravity models. The Gaussian Streaming Model (GSM) is shown to accurately reconstruct $\xi_2$  in modified gravity models and we employ the GSM, using a functional derivative approach, to analyze  the insensitivity of $\xi_2$ to the gravity model. Assuming linear theory, we show the void quadrupole to be an unbiased estimator of the redshift space growth rate parameter $\beta=f/b$ in the modified gravity theories.
\end{abstract}

\maketitle

\section{Introduction}

A century after Einstein dismissed the inclusion of a cosmological constant, $\Lambda$, in his field equations, such a term, equivalent to the addition of a non-zero vacuum energy, has since become the default explanation for the observed accelerated cosmic expansion \citep{Perlmutter:1998np, Riess:2004nr,Eisenstein:2005su, Percival:2007yw, Percival:2009xn, Kazin:2014qga, Spergel:2013tha, Ade:2013zuv, Ade:2015xua}. When the observationally inferred vacuum energy, $\Lambda_{obs}$, is compared against theoretical calculations from quantum field theory however, there is a discrepency of 120 orders of magnitude, leading to the cosmological constant problem. This incredible fine tuning has led to the consideration of alternative theoretical explanations for the observations.

One avenue of investigation is to induce deviations from General Relativity (GR) through the inclusion of new scalar degrees of freedom (see, for example, \citep{Clifton:2011jh}). These new gravitational degrees of freedom, referred to as the ``fifth force", typically lead to deviations from GR on cosmic scales $\sim 1/\Lambda_{obs}$. General Relativity is incredibly successful at predicting gravity on solar sytem scales. Hence, to remain viable, modified gravity theories must pass a plethora of strict earth and solar system scale tests \cite{Will_2006}. Theories which retain viability typically feature ``screening" mechanisms which suppress fifth-force modifications, and restore General Relativity, in solar system like environments.

In Hu-Sawicki $f(R)$ gravity \citep{Hu:2007nk}, the ``chameleon" mechanism \citep{Khoury:2003aq,Khoury:2003rn}  acts to increase the mass of the scalar field which mediates the additional fifth force in regions of high density, leading to a lack of propagation and suppression. Other alternative theories such as DGP gravity \citep{Dvali_2000} rely on the ``Vainshtein" mechanism \citep{VAINSHTEIN1972393}, which suppresses the fifth force whenever the derivatives of the additional scalar field grow large, such as inside and immediately surrounding a large overdense region. While these screening mechanisms suppress modifications to GR in high density environments, potential modifications to gravity would be expected to arise in cosmic voids, which are large underdense regions of the cosmic web. 

Cosmic voids have been observed in a wide range of cosmological surveys including photometric galaxy surveys (e.g. Dark Energy Survey \cite{S_nchez_2016}), spectroscopic galaxy surveys (e.g. SDSS/eBOSS \cite{Sutter_2014,Hawken_2020}) and CMB surveys (e.g. Planck \cite{Kov_cs_2022}). Voids can have a direct impact on weak gravitational lensing \citep{Krause:2012aq,Chantavat:2014gqa,Davies:2018jpm,Davies:2020udw,2020ApJ...890..168R}, redshift space distortions \citep{Hamaus:2015yza,Hamaus:2016wka,Cai:2016jek,2019MNRAS.483.3472N,Chuang:2016wqb,Sakuma:2017hfc,2019PhRvD.100b3504N,Correa:2020ddy,2020MNRAS.499.4140N}, CMB lensing \citep{Cai:2016rdk,Raghunathan:2019gyb,DES:2022uvb}, the integrated Sachs-Wolfe effect \citep{Nadathur:2011iu,2016ApJ...830L..19N}, and the kinetic Sunyaev-Zel’dovich effect \citep{Li:2020bsu} amongst others. Upcoming galaxy surveys such as Euclid \cite{Hamaus_2022}, the Dark Energy Spectroscopic Instrument (DESI) \citep{https://doi.org/10.48550/arxiv.1907.10688, 2016arXiv161100036D}, the Vera Rubin Large Synoptic Survey Telescope (LSST) \cite{2019ApJ...873..111I}, Roman Space Telescope \cite{https://doi.org/10.48550/arxiv.1503.03757}, and CMB surveys such as the Simons Observatory \cite{2019JCAP...02..056A}, and CMB-S4 \cite{2016arXiv161002743A} promise to add a wealth of cosmological data for void analyses.

Voids have been considered in the study of dark energy models \citep{Sheth2004AHO,Pisani:2015jha,Wojtak:2016brz,Adermann:2017izw,Contarini:2019qwf,Ceccarelli:2013rza,Ricciardelli:2014vga,Novosyadlyj:2016nnu,Massara:2018dqb, AragonCalvo:2012bd,Lambas:2015afa,Adermann_2018,2020PhRvL.124v1301N,Verza_2019} and as probes of massive neutrinos \citep{Schuster_2019,Massara_2015,Kreisch_2019,Bayer_2021}. The use of voids to identify potential modifications to gravity has been considered for a variety of models including  nDGP, $f(R)$, symmetron, and Galileon models \citep{Perico_2019,PhysRevD.104.023512,Li:2011pj,Zivick:2014uva,Perico:2019obq,Contarini:2020fdu,Padilla:2014hea,Cai:2014fma,Davies:2019yif,Falck:2017rvl,Paillas:2018wxs,Baker:2018mnu,Barreira:2015vra} 

In previous work \cite{PhysRevD.104.023512}, we showed that in $f(R)$ modified gravity scenarios, the fifth force leads to an enhancement to the void's coherent radial velocity profile -- the magnitude of which is dependent on the size of the void itself. While the radial velocity is not directly observable in itself, it does effect observable redshift space distortion statistics within the void environment.

In this paper we consider the effects induced by modifications to gravity on the void size function and on the redshift space void quadrupole moment to assess whether these effects might be observationally measurable. We also consider the application of the measured void quadrupole to recover a measure of the linear growth rate in the modified gravity theories, by constraining $\beta=f/b$, where $f$ is the cold dark matter (CDM) linear growth rate and $b$ is the bias of the observed tracers, as considered for GR in \cite{2019MNRAS.483.3472N}. 

This paper is structured as follows. Section \ref{sec:formalism} lays out the formalism. In Sec.~\ref{sec:results}, we present our main results centered around the measurable void statistics in modified theories of gravity. These findings are analyzed and contextualized in Sec.~\ref{sec:analysis}. They are also used to determine potential constraints on $\beta$ derived from the void quadrupole. In Sec.~\ref{sec:conc} we present the conclusions and implications of this work for future research.

\section{Formalism}
\label{sec:formalism}

In \ref{sec:MG}, we describe the two different modified gravity models, $f(R)$ and nDGP gravity, to be compared with the baseline $\Lambda \mathrm{CDM}$ model in this work. The choice of N-body simulations and application of Halo Occupation Distribution (HOD) is outlined in \ref{sec:sims}. The void finding and void stacking procedures are described in \ref{sec:voidFind} and \ref{sec:voidStack} respectively. Finally, the Gaussian streaming model is summarized in \ref{sec:GSM}.

\subsection{Modified Gravity Models}
\label{sec:MG}

\subsubsection{$f(R)$ gravity}

We modify the standard Einstein-Hilbert action \citep{Nojiri:2006gh} to instead take the form 
\begin{equation}
S_{f(R)}=\int d^{4}x \sqrt{-g}\left(\frac{1}{16\pi G}[R+f(R)] + \mathcal{L}_{m}(\psi_{i})\right),
\label{eq:ActionFR}
\end{equation}
where $f$ is some function of the Ricci scalar, $R$, and $\mathcal{L}_{m}(\psi_{i})$ is the standard model matter Lagrangian composed of fields $\psi_{i}$. In this paper, we consider the $f(R)$ form specified by Hu and Sawaki \citep{Hu:2007nk},
\begin{equation}
f(R) = -m^2 \frac{c_1 \left(R/m^2\right)^n}{c_2 \left(R/m^2\right)^n+1}.
\label{eq:HuSfR} 
\end{equation}
Here $m=H_0\sqrt{\Omega_{m0}}$ is a characteristic mass scale, with $H_{0}$ the Hubble constant, $\Omega_{m0}$ the fractional matter energy density today and free parameters $c_{1}$, $c_{2}$ and $n$ that are specified to fully define $f(R)$.

The modified field equations are obtained by varying the action with respect to the metric $g_{\mu \nu}$ 
\begin{equation}
G_{\mu \nu} + f_{R}R_{\mu \nu}-g_{\mu \nu}\left[\frac{1}{2}f(R)-\Box f_{R}\right]-\nabla_{\mu}\nabla_{\nu}f_{R} = 8 \pi G T_{\mu \nu}.
\label{eq:Gmunu}
\end{equation}
Here $G_{\mu \nu}$ is the Einstein Tensor $G_{\mu \nu} = R_{\mu \nu} - \frac{1}{2}g_{\mu \nu} R$, with $R_{\mu \nu}$ the Ricci tensor, $\Box=g^{\mu \nu}\nabla_{\nu}\nabla_{\mu}$ with $\nabla_{\mu}$ the standard covariant derivative with respect to $g_{\mu \nu}$, and  $f_R\equiv\frac{df(R)}{dR}$. 

In the limit of high curvature, $f_R$ is given by
\begin{equation}
f_R \simeq -n \frac{c_1}{c_2^2} \left(\frac{m^2}{R}\right)^{n+1}.
\label{eq:simplefR}
\end{equation}
In order to match the observed expansion history, $\bar{R}$ must remain unchanged from its $\Lambda CDM$ value, which gives
\begin{equation}
\bar{f}_{R} \simeq -\frac{n c_1}{c_{2}^{2}}\left[3\left(\frac{1}{a^{3}}+4\frac{\Omega_{\Lambda0}}{\Omega_{m0}}\right)\right]^{-(n+1)}.
\label{eq:fR0}
\end{equation}
This expansion history matching condition also fixes the ratio $\frac{c_{1}}{c_{2}} \approx 6 \frac{\Omega_{\Lambda 0}}{ \Omega_{m0}}$, as shown in \cite{Hu:2007nk}, leaving two free model parameters: $\frac{c_{1}}{c_{2}^2}$ and $n$. It is common in the literature to specify $\bar{f}_{R0}$, the background value of the field today $(a=1)$, rather than $\frac{c_{1}}{c_{2}^2}$. Smaller background field values lead to more screening of the fifth force and a smaller deviation from GR. In this work, we consider a $f(R)$ gravity model with a background field value of $10^{-5}$. We refer to the scenario as ``$F5$".

To consider perturbations, we assume a spatially flat, FRW metric using the Newtonian gauge with $\Phi$ and $\Psi$ denoting the gravitational potential and spatial curvature perturbations respectively, 
\begin{equation}
g_{\mu \nu} \mathrm{d}x^{\mu} \mathrm{d}x^{\nu}=a^2(\tau)\left[-(1+ 2 \Phi)\mathrm{d}\tau^2 + (1-2 \Psi) \gamma_{i j} \mathrm{d}x^i \mathrm{d}x^j\right],
\end{equation}
where $\tau$ is the conformal time, $a$ is the scale factor set to $a=1$ today and $\gamma_{i j}$ is the 3D metric on spatial slices of constant $\tau$.

In GR, the gravitational potential appearing in the metric which determines geodesics in the low energy limit is the Newtonian potential, $\Phi=\Phi_{N}$, which obeys the sub-horizon Poisson equation
\begin{equation}
\nabla^2 \Phi_{N} = 4 \pi G a^2 \delta\rho,
\label{eq:PhiN}
\end{equation}
where $\delta\rho = \rho - \bar{\rho}$, the deviation of the cold dark matter density from its mean background value. In $f(R)$ gravity, the new scalar field, $f_{R}$, acts to source a ``fifth force" through modifications to $\Phi$ so that instead of the total $\Phi$ satisfying \ref{eq:PhiN}, we have
\begin{equation}
\Phi = \Phi_{N} - \frac{1}{2} f_{R},
\label{eq:TotalfR}
\end{equation}
Where $\Phi_{N}$ still satisfies (\ref{eq:PhiN}), but the new scalar field $f_{R}$ satisfies a non-linear field equation given by 
\begin{equation}
\nabla^2 f_R = \frac{1}{3}a^2 \delta R(f_R)-\frac{8}{3}a^2 \pi G \delta \rho,
\label{eq:fieldfR}
\end{equation}
where $\delta R(f_R)= R(f_{R}) - \bar{R}$, and $R(f_{R})$ is solved for by inverting (\ref{eq:simplefR}). 

In $f(R)$ gravity, the chameleon mechanism is responsible for increasing the mass of the scalar $f_{R}$ in regions of high density thereby limiting its propagation. The flip side is that in regions of low density, this suppression shuts off and one may linearize the field equation for $f_{R}$ in order to gain intuition for the  fifth forces in void environments. Defining $\delta f_{R} = f_{R} - \bar{f}_{R}$ and linearizing (\ref{eq:fieldfR}) gives
\begin{equation}
\nabla^{2} f_{R} = a^{2} \mu^{2} \delta f_{R} - \frac{8}{3} \pi G a^{2} \delta \rho. 
\label{eq:linfR}
\end{equation}
Solving this equation for a $\delta$-function source yields the Yukawa potential, which features exponential suppression of $f_{R}$ far from the source. Thus, while the ratio of coupling constants says gravity is enhanced by at most $1/3$ in $f(R)$ over its GR value, we should expect the fifth force in $f(R)$ gravity to be short ranged relative to the Newtonian force. 

\subsubsection{Dvali-Gabadadze-Porrati (nDGP) gravity}

nDGP gravity \cite{Dvali_2000} assumes our 4D universe is confined to a 4D brane in a larger 5D spacetime with an action that includes both a 4D and a 5D term, 
\begin{eqnarray}
S_{nDGP}&=&\int d^{4}x \sqrt{-g}\left(\frac{1}{16\pi G}R + \mathcal{L}_{m}(\psi_{i})\right) \nonumber \\
&&+ \int d^{5}x \sqrt{-g^{(5)}}\frac{1}{16\pi G}R^{(5)}.
\label{eq:ActionDGP}
\end{eqnarray}
The first integral contains both the standard Einstein Hilbert $4D$ action as well as the matter fields, constrained to live on the $4D$ brane, while the second integral is the $5D$ Einstein Hilbert action. $g_{\mu\nu}$ is the induced 4D metric on the 4D brane consistent with the total $5D$ metric $g^{(5)}_{AB}$.

Varying the action with respect the full $5D$ metric yields the $5D$ modified Einstein equations, while applying the Israel Junction conditions across the 4D brane gives rise to the standard $4D$ Einstein equations on the braneworld plus a modification coming from the 5D bulk. A detailed analysis can be found in \cite{Koyama_2007}.  

On the brane, the $4D$ induced metric still takes the form (\ref{eq:Gmunu}), meaning in the low energy limit, particle geodesics are still determined by $\Phi$, where we now have 
\begin{equation}
    \Phi = \Phi_{N} + \frac{1}{2}\varphi
    \label{eq:totalnDGP}
\end{equation}
with $\varphi$ the ``brane bending mode" referred to in \cite{Koyama_2007}, and $\Phi_{N}$ the standard Newtonian potential obeying (\ref{eq:PhiN}). As can be seen from (\ref{eq:totalnDGP}), the fifth force is now sourced by $\varphi$, which obeys the non-linear field equation 
\begin{equation}
  \nabla^{2} \varphi + \frac{r_{c}^2}{3 \beta a^{2}}\left[(\nabla^{2} \varphi)^{2} - (\nabla^{i} \nabla^{j} \varphi)(\nabla_{i} \nabla_{j} \varphi) \right]=\frac{8 \pi G a^{2}}{3 \beta} \delta \rho.
  \label{eq:fieldVarphi}
\end{equation}
Here, $r_{c}$ is the crossover scale, defined by the ratio of the 5D and 4D newton constants \cite{Koyama_2007}
\begin{equation}
    r_{c}=\frac{1}{2} \frac{G^{(5)}}{G}
\end{equation}
and $\beta$ is a time dependent function given explicitly by \cite{10.1093/mnras/stab2817}
\begin{equation}
\beta = 1 + H_{0} r_{c} \frac{\Omega_{m0} a^{-3} + 2 \Omega_{\Lambda 0}}{\sqrt{\Omega_{m0} a^{-3} +  \Omega_{\Lambda 0}}}.
\end{equation}

The screening mechanism employed by nDGP gravity is the Vainshtein mechanism. That this mechanism causes suppression to the additional scalar degree of freedom scalar wherever its derivatives become large can be seen explicitly in  (\ref{eq:fieldVarphi}), in which the term in the square brackets, responsible for the screening, depends not on the value of $\varphi$ but on its derivatives. This is in contrast to the chameleon mechanism in $f(R)$ gravity which depends on the value of the additional scalar $f_{R}$ itself. 

Within the literature, different models of nDGP gravity are specified by the value of $r_{c} H_{0}$. Larger values of $r_{c}$ mean a stronger coupling to the screening term in square brackets and a smaller coupling to matter in (\ref{eq:fieldVarphi}). Hence, larger values of $r_{c} H_{0}$ lead to weaker modifications to gravity. In this work we consider the parameter value $r_{c} H_{0}=1$, which we refers to as ``N1", consistent with other work in the literature. 

Similar to our study of $f(R)$ gravity, we may linearize the nDGP field equation to gain intuition about the behavior of the fifth force in void environments. Linearizing (\ref{eq:fieldVarphi}) gives us 
\begin{equation}
  \nabla^{2} \varphi =\frac{8 \pi G a^{2}}{3 \beta} \delta \rho.
  \label{eq:linVarphi}
\end{equation}
The term in square brackets in  (\ref{eq:fieldVarphi}) has been dropped as it is $\mathcal{O}(\varphi^{2})$. At linear level, the fifth force in nDGP gravity features no additional screening, and thus should have long range solutions comparable to the Newtonian force within void environments. Comparing the matter coupling constants in  (\ref{eq:linVarphi}) and (\ref{eq:PhiN}) along with the extra factor of $1/2$ in (\ref{eq:totalnDGP}), with $H_{0} r_{c} = 1$, $\Omega_{M}\sim0.3$, and $\Omega_{\Lambda}\sim0.7$, gravity is increased by approximately $12\%$ in nDGP gravity when in the linear regime.

\subsection{Modified Gravity Simulations and HODs}
\label{sec:sims}

In this paper we use the MG-GLAM simulations described in \citep{Klypin:2018MNRAS.478.4602K.GLAM,Ruan_2022,Hernandez-Aguayo:2021arXiv211000566H.MGGLAM.DGP} to investigate nDGP and $f(R)$ gravity scenarios in comparison with a $\Lambda \mathrm{CDM}$ baseline. MG-GLAM is a particle-mesh code created to quickly simulate fully non-linear N-body simulations of modified gravity. MG-GLAM uses a multigrid relaxation technique to solve the non-linear field equations of these models, (\ref{eq:fieldfR}) and (\ref{eq:fieldVarphi}) for $f(R)$ and nDGP gravity respectively). 

The simulations consist of 100 realizations each for baseline GR with a $\Lambda \mathrm{CDM}$ cosmology, nDGP gravity with $H_{0} r_{c}=1$ (N1), and $f(R)$ gravity with $\vert\bar{f}_{R}\vert=10^{-5}$ (F5). The cosmological parameters are also the same as those described in \cite{Ruan_2022}, and are chosen to match the 2015 Planck cosmological parameters \cite{2016}. Explicitly, $\Omega_{m}=0.3089$, $h=0.6774$, $n_{s}=0.9667$, and $\sigma_{8}=0.8159$. The exact simulations used in this paper are larger than those presented in \cite{Ruan_2022}, although with the same mass and force resolutions. Each realization evolves $2048^3$ particles of identical mass $1.1\times10^{10} M_{\mathrm{sun}}/h$ in a periodic box of comoving size $L_{\mathrm{box}}=1024 \mathrm{Mpc/h}$, initialized at a redshift of $z_{\mathrm{initial}}=100$ with initial conditions generated using the Zel'dovich approximation \cite{1970A&A.....5...84Z}. All of our analysis is performed using the $z=0.5$ snapshot. We note that the $(1024 \mathrm{Mpc/h})^3$ simulation volume used at this redshift is indicative of that expected to be surveyed by DESI at $z=0.5 \pm 0.05$ \cite{Font_Ribera_2014}.

Particles are grouped into halos using the Bound Density Maxima (BDM) halo finder described in \citep{Klypin:1997astro.ph.12217K.BDM}. Each halo catalog is complete down to a minimum halo mass of $M_{\mathrm{Min}} \simeq 10^{12.5} M_{\mathrm{sun}}/h$, which is taken as the minimum halo mass for all analysis involving dark matter halos \footnote{Private communication}.

While different modified gravity models  will change the growth of dark matter large scale structure, as captured in the halo 1-point and 2-point functions  \citep{Falck_2015,Hern_ndez_Aguayo_2018}, it is the galaxies within dark matter halos, not the dark matter itself, that are observable. In order to get results which can be indicative of the statistics measured observationally with a galaxy survey, we must first augment the halo catalog with an appropriately tuned Halo Occupation Distribution (HOD) function to model how a realistic galaxy population is assigned to the halos \cite{Berlind_2003,Zheng_2005}. A HOD must be tuned to the simulations of each gravity theory separately so that the two-point galaxy correlation function matches that of a given, target observational dataset. We implement the HOD prescription laid out in \cite{Zheng_2007}, which is explicitly given by 
\bea
\left< N_{\mathrm{cen}}(M) \right> &=& \frac{1}{2}\left[ 1+\mathrm{erf}\left(\frac{\mathrm{log} M - \mathrm{log} M_{min}}{\sigma_{\mathrm{log}M}} \right)     \right]\\
\left<N_{\mathrm{sat}} (M) \right> &=& \left< N_{\mathrm{cen}} \right> \left(     \frac{M-M_{0}}{M_{1}}    \right)^{\alpha}
\eea
where $\left< N_{\mathrm{cen}}(M) \right>$ and $\left<N_{\mathrm{sat}} (M) \right>$ are respectively the number of central galaxies and satellite galaxies a halo of mass $M$ will hold on average. The model parameters are obtained by fitting to the simulation's 2-point function against observational survey data. Here, parameter values for $[M_{Min}, M_{0}, M_{1}, \sigma_{log\ M}, \alpha]$ for GR, F5, and N1 are taken from Table~II in \cite{alam2020testing}, with the GR parameter values representing the best fits to the BOSS CMASS DR9 dataset \cite{Manera_2012}, and modified gravity parameters tuned to match the resulting GR projected galaxy 2-point correlation function.

We compute the probability that a given halo of mass $M$ hosts a galaxy as 
\begin{equation}
    \left< N_{\mathrm{tot}}(M) \right> = \left< N_{\mathrm{cen}}(M) \right> + \left< N_{\mathrm{sat}}(M) \right>.
\end{equation}
If $\left< N_{\mathrm{tot}}(M) \right> \geq 1$, that halo is assumed to have at least one galaxy with certainty. If $\left< N_{\mathrm{tot}}(M) \right> < 1$, we assign a galaxy with probability $\left< N_{\mathrm{tot}}(M) \right>$.

\subsection{Void Finding Procedure}
\label{sec:voidFind}

We identify voids using the two different populations of tracers: the full set of halos and the subset of halos containing an HOD-identified mock galaxy.

Voids are identified in real space using the void finding package VIDE (Void IDentification and Examination toolkit) \citep{Sutter:2014haa}. VIDE implements ZOBOV (ZOnes Bordering On Voidness) \citep{Neyrinck:2007gy} which uses a Voronoi tessellation followed by a watershed algorithm to divide all of space into cells around each tracer, and then merge neighboring cells into ``zones" to identify depressions in the local matter density. All zones are identified as individual voids, with no additional merging of zones, as was also done in \cite{2019MNRAS.483.3472N}. For VIDE users, this means we select all ``bottom-level" voids, excluding all {\it parent} voids which are formed by joining regions with multiple {\it child} voids. This approach ensures that our definition of what constitutes a ``void" is only dependent upon the topology of the tracer density field \cite{2015MNRAS.454..889N}.

VIDE assigns each void an effective radius, $R_{\mathrm{eff}}$, such that a sphere of radius $R_{\mathrm{eff}}=\left(3V_{\mathrm{void}}/4\pi\right)^{1/3}$ would have equal comoving volume, $V_{void}$, to the void in question (which may not be spherical in itself).

Each void's location is specified by its {\it circumcenter}, defined as the center of the largest sphere entirely empty of tracers which can be circumscribed inside the void in real space \cite{Nadathur_2015A}. Note this is different from the void {\it macrocenter}, which is the VIDE default and defined as the volume weighted average position of all tracers within the void \cite{Sutter:2014haa}.

Void circumcenters are better indicators of the true density minimum of the void, and are more robust to the effects of redshift space distortions on void identification, as shown in \cite{2019MNRAS.482.2459N}. Note that the circumcenter of each void is empty of tracers by definition. This means that, consequently, we don't impose a central density threshold for void identification as VIDE does with macrocenters (typically requiring that a void must have a central density less than $0.2 \bar{n}_{sim}$).

\subsection{Void Stacking and Multipole Moments}
\label{sec:voidStack}

Although any particular void located in real space is not perfectly spherical, the lack of a preferred axis means that when many voids are averaged/stacked together, the resulting matter distribution will be highly spherically symmetric. When computed numerically, the average density contrast for stacked voids in real space is given by:
\begin{equation}
\delta^r(r) =\frac{N_{h}(r)}{N_v \bar{n}_{h} V(r) } - 1.
\label{eq:deltaRUnscaled}
\end{equation}
The superscript $r$ denotes real space, $n_{h}$ is the average number density of tracers (either of all halos or the subset of halos containing mock galaxies), $V(r)$ is the volume of the spherical shell ranging from $r-dr$ to $r+dr$, $N_{h}(r)$ is the total number of tracers in the shell, and $N_v$ is the total number of voids in the stack. 

In redshift space, spherical symmetry is no longer maintained. Using the distant observer approximation, tracer positions will shift from real space to redshift space according to 
\begin{equation}
\vec{s}=\vec{r}+\frac{\vec{v} \cdot \hat{l}}{\mathcal{H}},
\end{equation}
where ${\bf v}$ is the tracer velocity, $\hat{l}$ is the line of sight direction, commonly taken as  $\hat{x}$, $\hat{y}$, or $\hat{z}$, $\vec{s}$ is the position in redshift space, and $\vec{r}$ is the position in real space, and $\mathcal{H}=d\ln a/d\tau=a H(a)$ is the conformal time Hubble factor. 

The density contrast for stacked voids in redshift space is computed numerically using  
\begin{equation}
\delta^s (s,\mu_{s}) =\frac{N_{h}(s,\mu_{s})}{N_v \bar{n}_{h} V(s,\mu_{s}) } -1,
\label{eq:deltaSUnscaled}
\end{equation}
where tracers are now binned using both a radial coordinate relative to their void center, $s$, and angular coordinate relative to the line of sight (LOS), $\mu_{s}=\cos(\theta_{LOS})$. The superscript ``s" is used to denote the redshift space quantity. In both real space and redshift space, we use 50 equally separated radial bins of width $\Delta r = \Delta s = 2.4$ Mpc/h, and in redshift space, 100 equally separated angular bins for $\mu_{s}=\cos(\theta_{LOS})$ ranging from $[-1,1]$. $N_{h}(s,\mu_{s})$, and $V(s,\mu_{s})$ are defined similarly to their real space counterparts now with the inclusion of angular dependence. 

The void-galaxy multipole moments $\xi_{\ell}(s)$ for $\delta^{s}(s,\mu_{s})$ are defined as
\begin{equation}
\xi_\ell(s)=\frac{2 \ell +1}{2}\int_{-1}^{1} \delta^s(s,\mu_{s}) P_{2}(\mu_{s}) d \mu_{s},
\label{eq:multipoles}
\end{equation} 
where $P_{\ell}$ is the $\ell^{\mathrm{th}}$ Legendre polynomial. In this analysis, we focus our attention on the void quadrupole $\xi_{2}$, the first non-zero redshift space multipole moment induced entirely by the effects of redshift space distortions. In this theoretical analysis we consider the average $\xi_2$ signal over all 100 realizations and 3 independent line of sight directions.

We note that the commonality of void density profiles under a rescaling by each void's effective radius $\tilde{r}\equiv r/R_{\mathrm{eff}}$ has been used in the literature to motivate stacking voids using the ``rescaled" coordinate $\tilde{r}$ \citep{Hamaus_2014,Ricciardelli:2014vga}. In the main text of this paper, we present results for voids stacked without such a rescaling, however in Appendix \ref{app:rescaled} we also provide the accompanying results for voids stacked using rescaled coordinates.

\subsection{Gaussian Streaming Model}
\label{sec:GSM}

We employ the Gaussian Streaming Model (GSM) \cite{1980lssu.book.....P, Fisher_1995} to model the redshift space void quadrupole moment from the simulated real space data.

The coordinate change from real to redshift space coordinates,  $r$ and $s$ respectively, for transverse ($\perp$) and line of sight ($\p$) directions are given by
\begin{eqnarray}
    s_{\perp}&=&r_{\perp} \nonumber
    \\
    s_{{\p}}&=&r_{{\p}} + \frac{v_{r}(r) \mu_r}{\mathcal{H}} + \frac{v_{\p}}{\mathcal{H}}.
    \label{eq:randCoord}
\end{eqnarray}
Here $r=\sqrt{r_{\perp}^2 + r_{\p}^2}$, $\mu_{r}=r_{\p}/r$, $v_{r}$ is the coherent radial velocity flow in real space and $v_{\p}$ models random line of sight deviations around the coherent radial velocity.

The probability density function for $v_{\p}$, $\mathcal{P}$, is taken as a zero-mean Gaussian wholly specified by the dispersion in the line of sight velocity, $\sigma_{v_{\p}}(r,\mu_{r})$, which has both radial and angular dependence. In practice however, we find that the angular dependence is extremely weak compared to the radial dependence, with $\sigma_{v_{\p}}(r,\mu_{r})$ an increasing function of $\mu_{r}$, but with $\sigma_{v_{\p}}(r,\mu_{r}=1)$ only greater than $\sigma_{v_{\p}}(r,\mu_{r}=0)$ by on average $\sim 1\%$ when considering above median size voids in GR. Because of this, we neglect the angular dependence in $\sigma_{v_{\p}}$ and use $\sigma_{v_{\p}}(r) \simeq \sigma_{v_{\p}}(r,\mu_{r}=1) \equiv \sigma_{v_{r}}(r)$ , to quantify $\sigma_{v_{\p}}$ from the simulations in the remainder of this work. We find that this approximation does not change any of the results presented.

Explicitly, this means the probability density function for $v_{\p}$ is given by 
\begin{eqnarray}
  \mathcal{P}(v_{\p})  &=& \frac{1}{\sqrt{2 \pi} \sigma_{v_{\p}}(r)}  \exp\left(-\frac{v_{\p}^2}{2 \sigma_{v_{\p}}(r)^2} \right) ,
    \label{eq:Gaussian}
\end{eqnarray}
where the distribution $\mathcal{P}$ is always implicitly a function of $r$. The GSM allows us to write the redshift space density distribution in terms of the real space quantities as
\begin{eqnarray}
1+\delta^{s}(s,\mu_{s})&=& \mathcal{H} \int  d r_{{\p}}(1+\delta^{r}(r))\mathcal{P}\left(v_{\p}(r_{\p},s,\mu_s)\right) \ \ \
 \label{eq:GSMformal}
\end{eqnarray}
where, using (\ref{eq:randCoord}), $v_{\p}$ is  given by 
\begin{equation}
    v_\p=(s_{\p} - r_{\p}) \mathcal{H} - \mu_{r} v_{r}(r). \label{eq:vp}
\end{equation}
We note that (\ref{eq:GSMformal}) is equivalent to the expression used in \cite{2019MNRAS.483.3472N} (prior to expansion), where $v_{\p}$ is directly integrated over through the inclusion of the Jacobian from (\ref{eq:vp}). This is also consistent with integrating over a (non-zero mean) velocity variable  $v_{\p}\rightarrow v_{\p} + v_r\mu_r $ as in \citep{Hamaus:2016wka,Paillas_2021}.

\section{Results}
\label{sec:results}

In this section, results related to the impact of the HOD on the void size function are discussed in \ref{sec:HOD}. The dynamical properties of the HOD-identified void populations are described in \ref{sec:voidProperties}. Section ~\ref{sec:quad} presents the redshift space quadrupole results across each theory of gravity.

\subsection{Void Size Function}
\label{sec:HOD}

While halos provide a useful mechanism for studying dark matter properties, we also need to consider how surveys will sample the halos with the galaxy tracers they observe. We connect the two  by considering a halo occupation function (HOD) that assigns mock galaxy tracers to dark matter halos in such a way as to reproduce a target (would-be observed) two-point galaxy correlation function, as described in  \ref{sec:sims}. 

In this section we compare the void count statistics for voids identified from halos and HOD derived mock galaxies in GR and the modified gravity models. We then assess how the imposition of a HOD modifies the predicted size and number density of identified voids, which will be important for the subsequent analysis.

\begin{table}[t!]
\begin{tabular}{|c|c|c|c|c|}
\hline
\multirow{3}{*}{Model}&  \multicolumn{2}{c|}{Halo-identified} &  \multicolumn{2}{c|}{HOD-identified} 
\\ \cline{2-5}
 & \multirow{2}{*}{\# Voids} & Median R$_{\mathrm{eff}}$  & \multirow{2}{*}{\# Voids}  & Median R$_{\mathrm{eff}}$ 
 \\ & & (Mpc/h) & &(Mpc/h) 
 \\ \hline
GR & 9,034 $\pm$ 76 & 24.7 $\pm$  0.1  &3,811 $\pm$  50  &   34.9 $\pm$  0.2  
\\ \hline
F5 & 10,206 $\pm$ 74 & 23.7 $\pm$ 0.1 &3,775 $\pm$ 46   & 35.0 $\pm$  0.2   
\\ \hline
N1 & 9,118 $\pm$  73 & 24.6 $\pm$ 0.1  & 3,776 $\pm$  52   & 35.0 $\pm$ 0.2 
\\ \hline
\end{tabular}
\caption{ Comparison of the mean number of voids per realization, and their median effective radius, $R_{\mathrm{eff}}$, for voids identified using halos [left] and HOD mock galaxies [right] as tracers, for each of the GR, F5 and N1 models. The 1$\sigma$ statistical variations in the average values in  one realization are also given.}
\label{tab:voidStats}
\end{table}

\begin{figure*}[!t]
\includegraphics[width=1.0\linewidth]{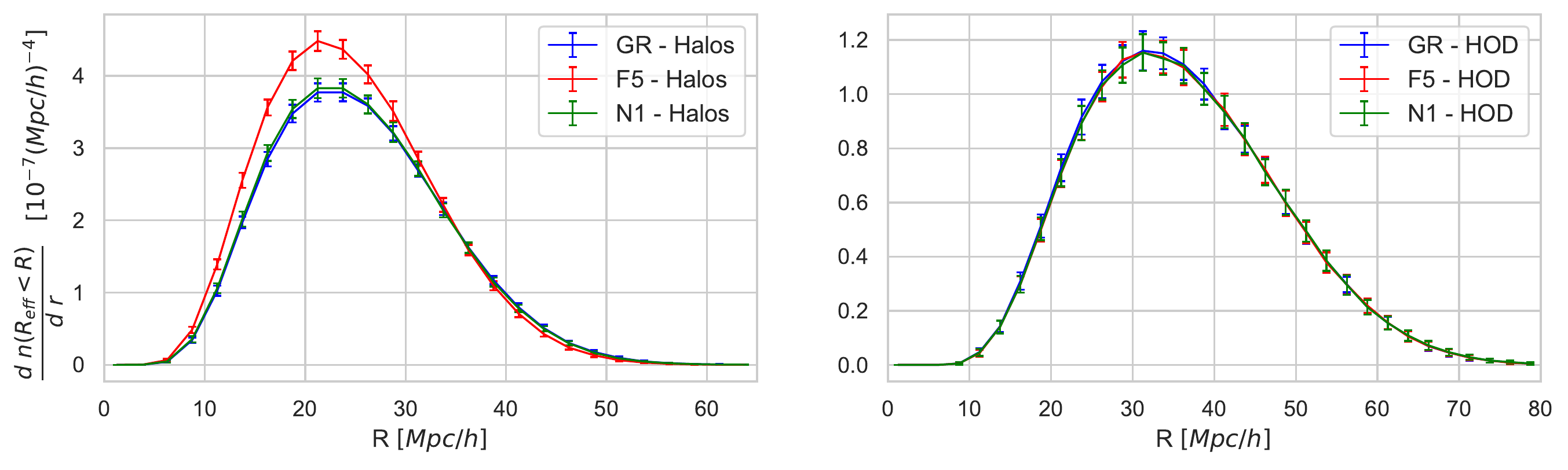}
\caption{The average void size function at $z=0.5$ for GR [blue], and $f(R)$ F5 [red] and nDGP N1 [green] in each model. Left: Voids are identified using halos with a minimum mass $10^{12.5} M_{\mathrm{sun}}/h$. Right: Voids are identified using a HOD tuned to ensure consistency in the galaxy 2-point correlation function. Error bars represent statistical uncertainties on one realization.}
\label{fig:VSF}
\end{figure*}

The total void population counts and average sizes are summarized in Table~\ref{tab:voidStats} for both halo and HOD-identified voids.  When halos, rather than HOD mock galaxies, are used we have more voids with a smaller median size, and larger differences between each theory of gravity. When voids are identified using HOD mock galaxies, the number of voids decreases by $\sim$60\%, while the median size of the void populations increases by about 40\%. The results in Table~\ref{tab:voidStats} show how using HOD mock galaxies as tracers eliminates differences in the median void size across theories of gravity, and brings the total number of voids per realization back into agreement across all three theories.

Figure~\ref{fig:VSF} provides more detail by showing the effect that tracer selection in the void identification process has on the resulting void size function (VSF), which we define as the number density of voids as a function of their effective radius. 

Relative to the statistical uncertainties, the void size function for voids identified from halos can clearly distinguish between GR and F5, but not between N1 and GR. In $f(R)$ theories structure growth is enhanced relative to GR, leading to a larger number of halos above the minimum mass cutoff. The total number of voids identified with VIDE watershed algorithm is heavily dependent on the number of tracers used to identify voids \cite{Jennings_2013}. Given a set of tracers on which a watershed algorithm has been run, adding additional tracers can only ever increase, or leave unchanged, the number of catchment basins identified. As such, the relative increase of $13\%$ in the number of voids identified in F5 relative to GR follows naturally from the $12\%$ increase in the number of halos in the F5 simulations.  N1, by contrast, has a 2\% fractional increase in number of halos relative to GR, and this gives rise to a $1\%$ increase in the number of voids, indistinguishable from GR when considered relative to the estimated errors.

When we switch from halos as tracers to HOD-populated mock galaxies, Fig.~\ref{fig:VSF} shows that the differences between GR and F5 disappear; the VSF for GR, N1 and F5 generally align well within the statistical uncertainties for the GR sample. This dramatic change in the HOD-derived void number function relative to that derived from halos highlights the need to take care in the void identification process.

\begin{figure*}[!t]
\includegraphics[width=1.0\linewidth]{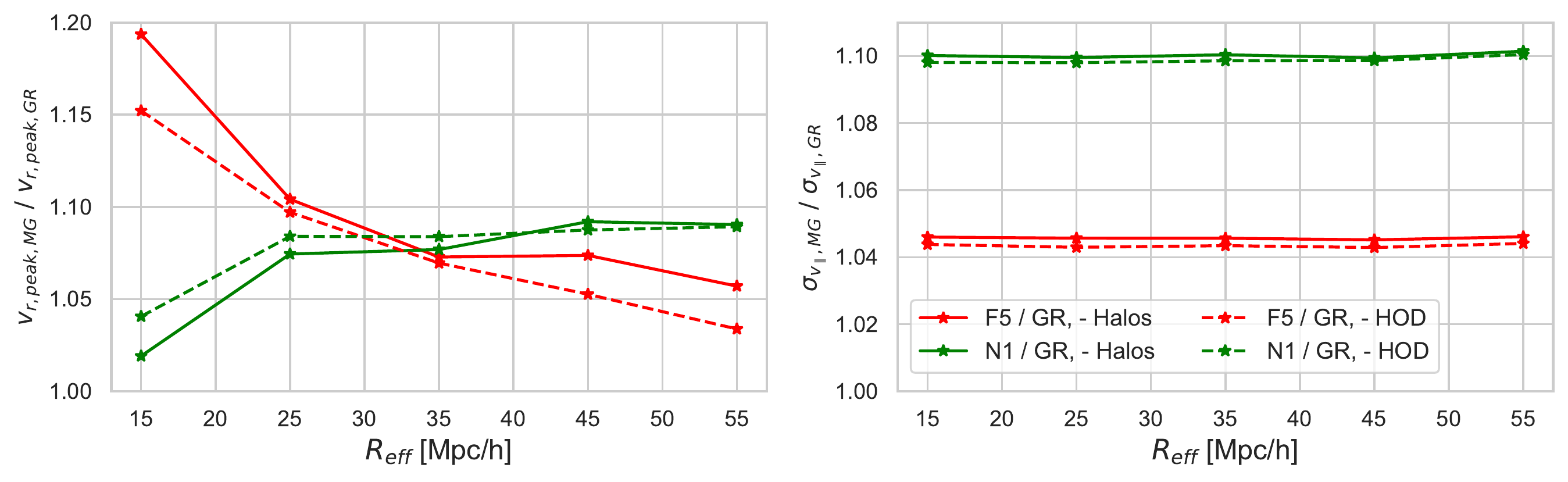}
\caption{Left: Ratio of the peak radial velocity in F5 [red] and N1 [green] to that in GR as a function of void effective radius, $R_{\mathrm{eff}}$, for voids identified from halos [solid] and from halos containing mock-galaxies identified by the HOD [dashed]. Right: Ratio of the average asymptotic value of the velocity dispersion, at $r\gtrsim 40$ Mpc/h, as a function of $R_{\mathrm{eff}}$.}
\label{fig:ratios}
\end{figure*}

\subsection{Void Density and Velocity Properties}
\label{sec:voidProperties}

Given that the void size function is equalized between GR and both N1 and F5 when voids are identified using HOD mock galaxies as opposed to halos, it is natural to ask how other void properties might be affected in each modified gravity model relative to their GR values. Due to the differences in the fifth force screening mechanism between each modified gravity theory (chameleon in $f(R)$, Vainshtein for nDGP), the application of an HOD has the potential to effect void properties in F5 and N1 in distinctly different ways. 

Figure ~\ref{fig:ratios} shows the effects of the modifications to gravity, as well as the effects of tracer selection in the void identification process (halo or mock galaxy), on the radial velocity and velocity dispersion as a function of void size. 

The ratio of the peak radial velocities,  $v_{r,peak,MG} / v_{r,peak,GR}$, between F5 or N1 with respect to the GR value demonstrate the scale dependent nature of each of the respective screening mechanisms. As shown in \cite{PhysRevD.104.023512}, the screening mechanism in $f(R)$ gravity is size dependent, and the magnitude of the fifth force to that of the Newtonian force for similarly shaped void density profiles is a decreasing function of $R_{\mathrm{eff}}$. By contrast, when linearized, the field equation for the fifth force in N1 (\ref{eq:linVarphi}) shows no scale dependence.  Accordingly, the peak velocity ratio $v_{r,peak,N1} / v_{r,peak,GR}$ is much more constant with respect to $R_{\mathrm{eff}}$ compared to the F5 case. 

The use of an HOD also dampens the F5 radial velocity profiles to a much larger degree than for N1. The screening mechanism in $f(R)$ gravity is known to be environmentally dependent \citep{Falck_2015, Shi_2017}, whereas the use of an HOD preferentially selects more massive halos to receive a mock galaxy and thus be used as tracers. Hence, we attribute the reduction in $v_{r,peak,F5} / v_{r,peak,GR}$ in the HOD selected sample relative to the halo selected one, seen in Figure~\ref{fig:ratios}, to the preferential selection of more massive (and therefore more screened) halos, where the effect of the fifth force is reduced in F5. 

The effect of the HOD on the N1 velocities is much less pronounced and provides neither a consistent increase nor decrease to $v_{r,peak,N1} / v_{r,peak,GR}$. In nDGP gravity, the nature of the Vainshtein screening mechanism is fundamentally different than that of the Chameleon mechanism in $f(R)$ gravity. The Vainshtein mechanism heavily screens short wavelengh $\varphi$ modes, such as those generated within halo environments, while leaving long wavelength modes to remain mostly unscreened. For the N1 parameter value, almost all halos are already screened \citep{Falck_2015,Falck_2014} -- consistant with the previously mentioned modest $2\%$ fractional increase in the number of halos in N1 over GR. This means that the preferential selection of more massive halos through an HOD provides no additional suppression to the N1 velocities. Despite all halos in N1 appearing as screened, the derivative shift symmetry present in ( \ref{eq:fieldVarphi}), $\partial_{\nu}\varphi \rightarrow \partial_{\nu}\varphi + c_{\nu}$, enables long wavelength $\varphi$ modes from distant sources in superposition with the $\varphi$ profile sourced by a screened halo to together solve (\ref{eq:fieldVarphi}). This in turn allows these screened halos to move in response to the long wavelength $\varphi$ modes as if they were unscreened point particles, which allows for increases to $v_{r}$ in N1 despite each tracer being screened \cite{Hui_2009}. This is also what allows the use of the linearized field  (\ref{eq:linVarphi}) within void environments, as all relevant $\varphi$ modes are long wavelength, sourced by the void environment itself or from distant halos.

Figure~\ref{fig:ratios} also shows the relative change to the asymptotic value of $\sigma_{v_{\p},N1}$ or $\sigma_{v_{\p},F5}$ with respect to the GR value as a function of void size. In both N1 or F5, and using either halos or mock galaxies to identify voids, there is no scale dependence in this value. 

The greater velocity dispersion in modified gravity scenarios relative to GR is well documented (e.g. \cite{Hellwing_2014}), and can be traced back to the action of the fifth force increasing peculiar velocities. While the void environment is the driving factor in determining the average $v_{r}$, the same is not true of the velocity dispersion. Immediately outside the void, regardless of void size,  we find that $\sigma_{v_{\p}}$ returns to its background value across the simulation as a whole, indicating that void environments themselves do little to determine the asymptotic value of $\sigma_{v_{\p}}$.

\begin{figure*}[!t] 
\includegraphics[width=1.0\linewidth]{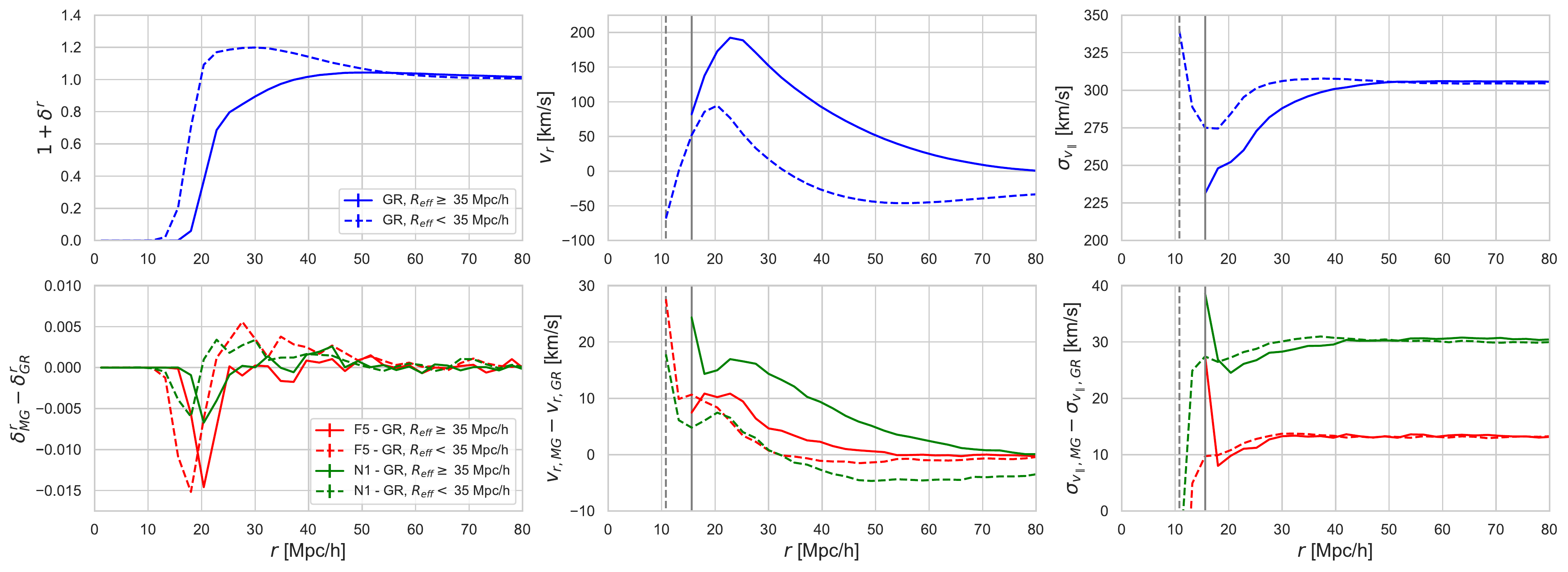}
\caption{Top Row: Void density [left], radial velocity [center] and velocity dispersion [right] profiles in GR for voids identified in halos with HOD-identified tracers. The void population is split into those with void size above [solid] and below [dashed] $R_{\mathrm{eff}}$=35Mpc/h, the median void size found in each of the three theories. Gray lines indicate the first radial bin below which no tracers are found. Bottom Row: Differences between F5 [red] and N1 [green] quantities with respect to those in GR.}
\label{fig:densVelsSigmas}
\end{figure*}

Given the scale dependent nature of $v_{r}$ in each modified gravity theory, it is informative to investigate behavior in two groups distinguished by size. We break the sample into two populations of smaller and larger voids, respectively of sizes below and above the median size, $R_{\mathrm{eff}} =35\mathrm{Mpc/h}$. 

Figure~\ref{fig:densVelsSigmas} shows the real space density profile, $\delta^{r}$, the void radial velocity profile, $v_{r}$, and the velocity dispersion profile, $\sigma_{v_{\p}}$, for the large and small void populations, for mock galaxy identified voids in GR, while the bottom row displays the change in each of these quantities in F5 or N1 over the respective GR value.

The interiors of the voids are extremely rare with $1+\delta_r\approx0$. At the void edge, the density rapidly increases, with a mild overdensity at the void edge, which is more pronounced in small voids than large. The modified gravity models do not significantly change the density profile.

The radial velocity, $v_{r}$, is an increasing function inside the void interior before reaching its peak value just around the void edge, where the density contrast increases rapidly. The radial velocity from the large void sample is constantly larger in magnitude than that of the small voids, in accordance with linear theory. We find that both F5 and N1 increase the size of the velocity peak relative to GR, with both providing comparable increases in the small void population, but N1 providing a larger and more spatially extended increase within the large voids.

The relative effects of the modified models can be understood through the linearized field equations. For the $f(R)$ field (\ref{eq:linfR}) yields the Yukawa potential, and the resulting fifth force is short-ranged in comparison with Newtonian gravity. In contrast, the linearized N1 field equation (\ref{eq:linVarphi}) features no screening term and is thus longer ranged than its F5 counterpart. This leads to a modification to the radial velocity profile, $\Delta v_{r}$, which remains non-zero in N1 gravity over a much larger spatial extent compared to F5. This phenomenon can be seen in Fig.~\ref{fig:densVelsSigmas} for both small and large void populations alike. The $R_{\mathrm{eff}}$ dependent enhancement to $v_{r}$ seen in Fig.~\ref{fig:ratios} can also be seen in Fig.~\ref{fig:densVelsSigmas}. In F5, the value of $\Delta v_{r} / v_{r,GR}$ is largest in the small void population, while this quantity is much more similar in both void populations for N1. 

While the qualitative changes to the radial velocities are quite distinct, the changes to the velocity dispersion induced by the modified gravity models are similar in F5 and N1. While $\Delta v_{r}$ is largest within the void interior and the exact shape is sensitive to the intricacies of the particular model, increases in $\Delta\sigma_{v_{\p}}$ relative to GR are largely independent of the distance from the void, with $\sigma_{v_{\p}}$ tending towards a value determined by the large scale properties of the simulated density field, rather than the void itself.

In Fig.~\ref{fig:densVelsSigmas}, $(1+\delta^{r})$ is at most $\sim 1\%$ different between GR and either modified theory in both void populations shown while peak values of $v_{r}$ are increased at roughly the 5-10\% level over the GR values depending on the void population and theory of gravity considered. In the large void population, $v_{r,peak}$ is increased by $9\%$ in N1 and $6\%$ in F5, while in the small void population, these numbers change to $8\%$ in N1 and $9\%$ in F5 respectively. For $\sigma_{v_{\p}}$, the asymptotic values increase by $10\%$ in N1 and $4\%$ in F5 independent of void population. In the following section we discuss the implications of these differences on the void quadrupole statistic.

\begin{figure*}[!t]
\includegraphics[width=1.0\linewidth]{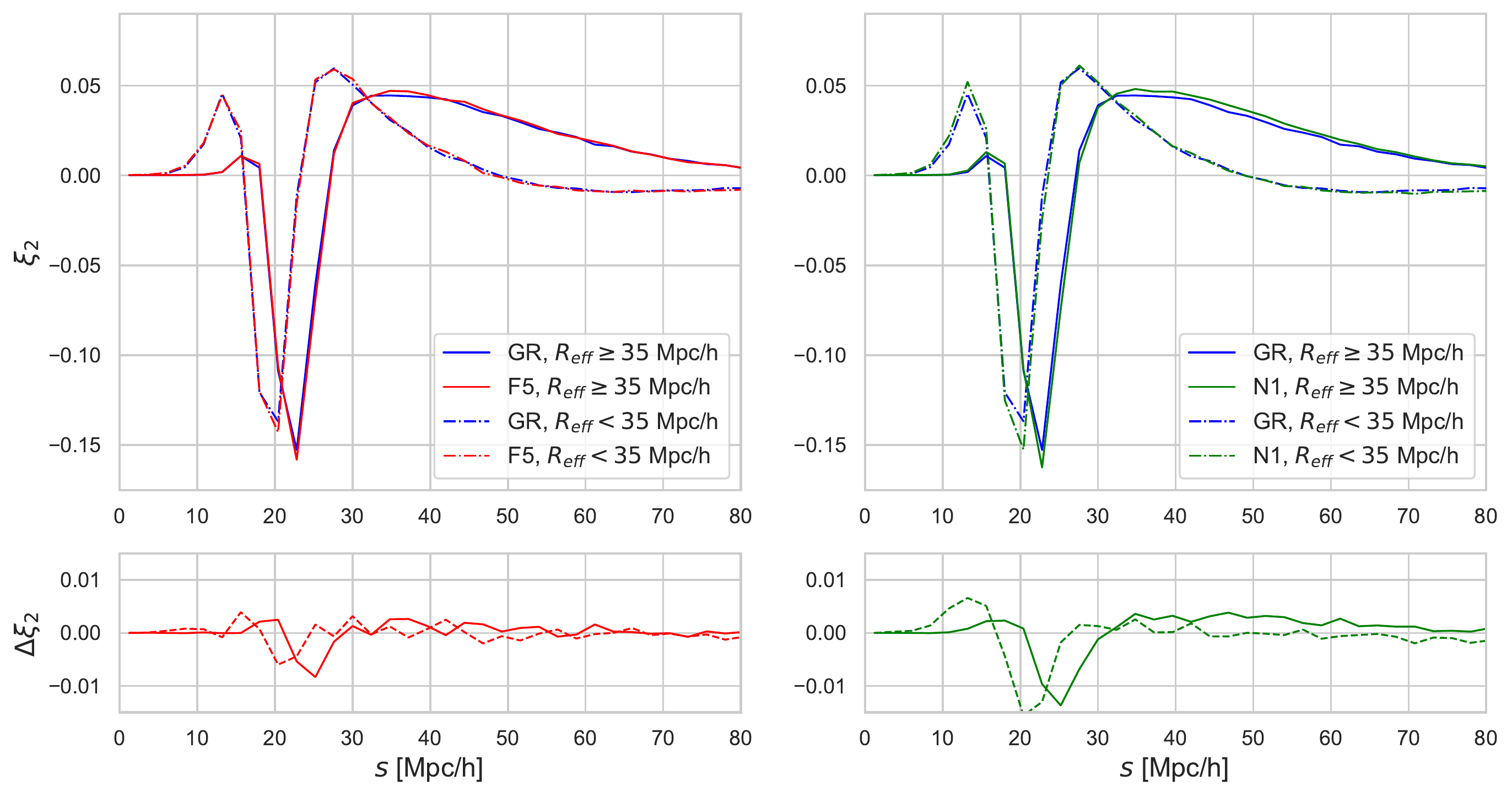}
\caption{Top Row: The void quadrupole moment $\xi_{2}$  at $z=0.5$ for GR [blue] compared against F5 [left, red] and N1 [right, green] in voids both below [dashed] and above [solid] the median void size. Bottom Row: Difference in $\xi_{2}$ between GR and each modified gravity theory in small [dashed] and large [solid] voids.}
\label{fig:quadData}
\end{figure*}

\subsection{Void Quadrupole Moments}
\label{sec:quad}
In real space, stacked voids are spherically symmetric, and therefore the only non-zero multipole moment will be the monopole $\xi_{0}$, equal to the real space density contrast $\delta^{r}(r)$. For stacked voids in redshift space, RSD effects under the distant observer approximation break this spherical symmetry along the line of sight, while preserving reflection symmetry across the plane which passes through the center of the void orthogonal to the LOS direction. The consequence of this reflection symmetry is that $\delta^{s}(s,\mu_{s}) = \delta^{s}(s,-\mu_{s})$, which sets all redshift space $\xi_{\ell}$ with odd $\ell$ identically equal to 0. Thus, the quadrupole, $\xi_{2}(s)$, is the first non-zero multipole moment induced entirely by redshift space distortions and will be the the focus of the remainder of this work.

Given two of the three functions used by the GSM to predict $\delta^{s}(s,\mu_{s})$ and $\xi_{2}$ see changes under modifications to gravity at roughly the $10\%$ level, as discussed in the previous section, one might intuit a similar level of change in $\xi_{2}$ for at least some of the void populations and modified gravity scenarios considered.

In Fig.~\ref{fig:quadData} we present $\xi_{2}(s)$ for HOD-identified voids in GR, F5 and N1 in both the large and small void populations (defined with respect to the median void size, $R_{\mathrm{eff},median}\simeq$35 Mpc/h). 

For N1 gravity,  given Fig.~\ref{fig:ratios}, we  expect the largest deviation between $\xi_{2,N1}$ and $\xi_{2,GR}$ to occur in large voids. Examining the population of larger voids, we find $\xi_{2,N1}$ is on average $8\%$ larger than $\xi_{2,GR}$ for $s\sim $35-65 Mpc/h. 

In the small void population, we see little to no difference between $\xi_{2}$ for  N1 and GR, except for a very localized change in the minimum at the void edge.

Comparing F5 versus GR, we find  $\xi_{2,F5}$ and $\xi_{2,GR}$ are strikingly similar in both the void populations except for some variation tightly located at the void edge. Although this might have been expected for the large voids, given the scale dependent screening produces the largest $\Delta v_{r} / v_{r}$ in small voids (as in Fig. ~\ref{fig:ratios}), the fact that there is little difference between $\xi_{2,F5}$ and $\xi_{2,GR}$ in the below median size void population doesn't follow that simple intuition.

\section{Analysis}
\label{sec:analysis}

\subsection{Gaussian Streaming Model Predictions in Modified Gravity}
\label{sec:GSMMG}

We seek to understand why differences on the order $10\%$ in the velocity variables, $v_{r}$ and $\sigma_{v_{\p}}$ (which both feature drastically different spatial profiles), don't induce comparable variations in $\xi_{2}$ between GR and the modified theories. To do this we employ the Gaussian Streaming Model which takes as inputs $\delta^{r}(r)$, $v_{r}(r)$, and $\sigma_{v_{\p}}(r)$ and returns a prediction for  $\xi_{2}(s)$.

\begin{figure*}[!t]
\includegraphics[width=1.0\linewidth]{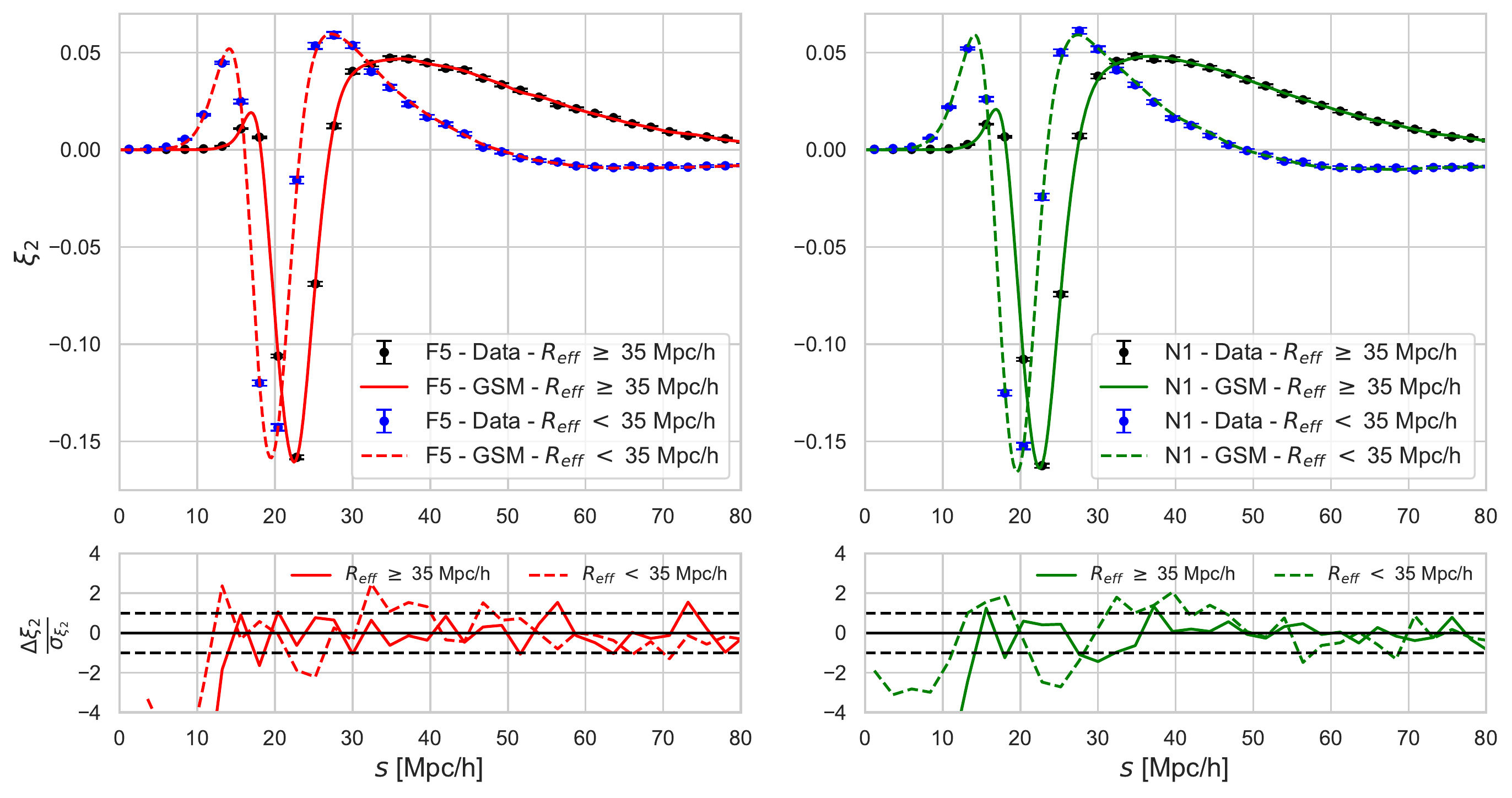}
\caption{Top: Theoretical value of $\xi_{2}$ in voids both above [dashed] and below [solid] the median void size in F5 [left, red] and N1 [right, green] calculated using the Gaussian Streaming Model (GSM) compared to average $\xi_{2}$ from the simulations. Bottom: Difference between the GSM-derived quadrupole and that from the simulated data, $\Delta\xi_2=\xi_{2,GSM}-\xi_{2,Data}$, with respect to the statistical uncertainties for the model, $\sigma_{\xi_{2}}$.}

\label{fig:quadThy}
\end{figure*}

We first assess how well the GSM performs in each theory of gravity. Figure~\ref{fig:quadThy} compares the simulation-derived quadrupoles with the theoretical GSM predictions for N1 and F5 for both the large (above-median radius) and small (below-median) populations of voids identified with halos containing HOD-identified galaxies.  Note, to assess the accuracy and precision of the GSM prediction for each model we compare the differences of the mean values relative to the errors on the mean signal from the full 100 realizations.

When taking the values of $\delta^{r}$, $v_{r}$, and $\sigma_{v_{\p}}$ directly from the simulations, the GSM performs exceptionally well in all three gravity scenarios (GR not shown), and for both void size groups.

In both the small and large void population, at radial distances exceeding $15 \mathrm{Mpc/h}$, the GSM rarely exceeds statistical uncertainties in the mean estimated from the 100 realizations. As such, the GSM model is shown to provide a robust method with which to model the contributions to the quadrupole.

\subsection{Dissecting the quadrupole}
\label{sec:dissect}

Given the GSM allows us to accurately model the quadrupole, we now use it to dissect and explain why the differences between $\xi_{2}$ in GR and modified gravity are so much smaller than the differences in the contributing $v_{r}$ and $\sigma_{v_{\p}}$ might suggest. 

As a mathematical object, the GSM is a functional, which we will denote with $F$. For each model of gravity, at a given redshift space radial coordinate, $s$, it takes the three void density and velocity functions and returns the quadrupole $F \left(s,\delta^{r}, v_{r}, \sigma_{v_{\p}} \right) \rightarrow \xi_{2}(s)$. The functional picture of the GSM allows us to mix and match quantities from different gravitational scenarios to predict the hypothetical $\xi_{2}$. For example, we could compute $\mathrm{F}\left(s,\delta^{r} \vert_{GR}, v_{r} \vert_{F5}, \sigma_{v_{\p}} \vert_{GR} \right)$ in order to isolate the effect that F5's increased radial velocity has on $\xi_{2}$, while holding the density and velocity dispersion fixed at their GR values. These expressions in turn, where one function takes on its F5 or N1 value and the others remain fixed at their GR values, can be calculated by taking the appropriate functional derivative of the GSM, and integrating against the change in the corresponding quantity.

The total change in $\xi_{2,F5}(s)$ or $\xi_{2,N1}(s)$ relative to $\xi_{2,GR}(s)$ can be approximated to a high degree of accuracy as the sum of the individual first-order changes to $\xi_{2}$ induced by the individual changes to $\delta^{r}$, $v_{r}$, and $\sigma_{v_{\p}}$ in the modified gravity (MG) theories,
\bea
     \xi_{2,MG} - \xi_{2,GR} \equiv \Delta\xi_2(s) 
     &=& \Delta_{\delta^r}\xi_{2}  
    + \Delta_{v_{r}}\xi_{2}
    + \Delta_{\sigma_{v_{\p}}}\xi_{2}.
    \label{eq:SumLinearChanges}
\eea
where $\Delta_{x}\xi_{2} \equiv \left(\mathrm{F} \left(x\vert_{MG},... \right)- \mathrm{F}\vert_{GR} \right)$ for $x=\delta^{r}(r)$, $v_{r}(r)$, or $\sigma_{v_{\p}}(r)$. Each of the terms in (\ref{eq:SumLinearChanges}) can be written as,
\bea
    \Delta_{x}\xi_{2}(s) &=& \int dr \frac{\delta F}{\delta x}\Delta x
    \nonumber \\
     &=&\frac{5}{4}\int^{1}_{-1} d \mu_{s}\  (3\mu_{s}^2 - 1)  \Delta_{x}\delta^{s}(s,\mu_{s}) \label{eq:generalFD}
\eea
Here $\Delta x$ indicates the difference in values of variable $x$ between the MG and GR models and $\Delta_{x}\delta^{s}$ denotes the change induced in $\delta^{s}$ induced by the change in $x$. Suppressing the arguments of $v_{\p}(r_{\p},s,\mu_{s})$ for brevity, these functions are given by 
 \bea
   \Delta_{\delta^{r}}\delta^{s} &=& \mathcal{H}  \int d r_{\p}\mathcal{P}\left(v_{\p} \right)\Delta \delta^{r}(r) \label{eq:delFD} 
   \\
   \Delta_{v_{r}}\delta^{s}&=& \mathcal{H} \int d r_{\p}\frac{\mu_{r} v_{\p} }{\sigma_{v_{\p}}(r)^2} \mathcal{P}\left(v_{\p} \right)(1+\delta^{r}(r))
     \Delta v_{r}(r) \label{eq:velFD} 
 \\ 
 \Delta_{\sigma_{v_{\p}}}\delta^{s} &=& \mathcal{H}\int \frac{d r_{\p}}{\sigma_{v_{\p}}(r)}\mathcal{P}\left(v_{\p} \right) \left(\frac{v_{\p}^2}{\sigma_{v_{\p}}(r)^2} - 1\right) (1+\delta^{r}(r)) \Delta \sigma_{v_{\p}} \nonumber\\ \label{eq:sigmaFD} &&
 \eea

\begin{figure*}[!t]
\includegraphics[width=1.0\linewidth]{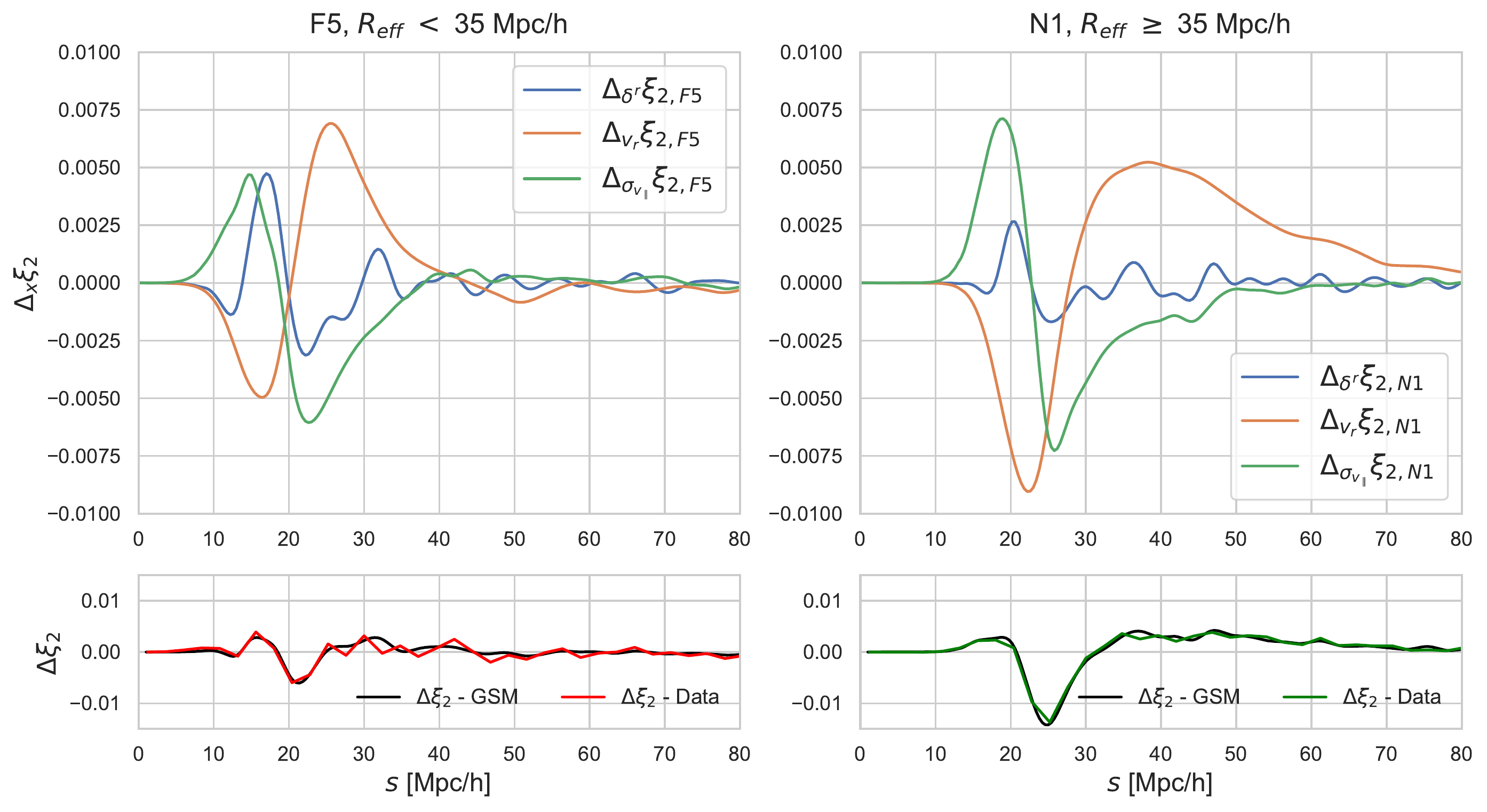}
\caption{Two scenarios are considered: [Left] Below median size voids in F5 and [Right] above median size voids in N1. Top Row: The changes to $\xi_{2}$ relative to GR induced independently by $\Delta \delta^{r}$ [blue], $\Delta v_{r}$ [orange], and $\Delta \sigma_{v_{\parallel}}$ [green] as calculated using GSM functional derivatives. Bottom Row: Comparison of the differences,  $\Delta\xi_{2}$, between the modified gravity model and GR obtained from the simulated data in F5 [left,red] and N1 [right, green] and the summed GSM integrated functional derivatives [black].}

\label{fig:funcDeriv}
\end{figure*}

Figure~\ref{fig:funcDeriv} shows the changes induced in $\xi_{2}$ independently by changes to $\delta^{r}$, $v_{r}$ and $\sigma_{v_{\p}}$ respectively, as calculated using the functional derivative approach and compares them to the total changes in $\xi_{2}$ from the simulations. The figures show results for the small void population in F5 and the large void population in N1 motivated by Fig.~\ref{fig:ratios}, in which it is small voids in F5 and large voids in N1 which would receive the greatest increases in their radial velocity profiles and therefore provide the most interesting case studies. The total $\Delta\xi_{2}$, as calculated from the simulation data directly, is shown to be accurately reproduced by the sum of the individual changes as calculated using the functional derivative approach described in (\ref{eq:SumLinearChanges}). As a result, we  use this approach to examine the individual terms to understand their respective roles in inducing changes to $\xi_{2}$ in the modified gravity scenarios.

To get an intuitive understanding of the individual changes shown in Fig.~\ref{fig:funcDeriv}, we can consider the contributions from the line of sight,  $\mu_{s}=\mu_{r}=\pm1$. This is where the bulk of the contribution to both $\Delta_{v_{r}}\xi_{2}$ and $\Delta_{\sigma_{v_{\p}}}\xi_{2}$ are derived, which can be understood by noting that $\mu_{s}=\pm1$ is where $P_{2}(\mu_{s})\propto (3 \mu_{s}^2 - 1)$ is maximized, and also where $v_{r}$ may have the greatest impact on the real to redshift space coordinate change in accordance with (\ref{eq:randCoord}).

Figure~\ref{fig:funcDeriv} shows that changes induced by $\Delta v_{r}$ provide the dominant contribution to $\Delta \xi_{2}$ outside the void in both small F5 voids and large N1 voids. Physically, the positive increases induced by the MG models to the outflowing radial velocity $v_{r}$ shift matter further away from the void under redshift space distortions. This causes a negative spike in $\Delta_{v_{r}}\delta^{s}(s,\mu_{s}=1)$ (and therefore $\Delta_{v_{r}}\xi_2(s)$)
closer to the void (where tracers streamed to without the increase) and a positive spike at larger $s$ (where tracers stream to now).

The negative and positive peaks in $\Delta_{v_{r}}\xi_2(s)$ are roughly localized to the radial positions with coincident large positive $\Delta v_{r}(r)$ and non-zero $(1+\delta^{r})$. In F5, due to the short range of the fifth force, $\Delta v_{r}$ is also short-ranged, and does not extend far beyond the void interior. Hence, $\Delta_{v_{r}} \xi_2(s)$ is also spatially limited to the void edge region when considering F5 modifications. In N1, the fifth force is longer ranged, and thus so is $\Delta v_{r}$ - extending well beyond the void edge out to almost $80 \mathrm{Mpc/h}$ as shown in Figure~\ref{fig:densVelsSigmas}. This causes $\Delta_{v_{r}} \xi_2(s)$ to decouple from the void edge region and to have a much longer spatial extent in the N1 voids considered, as can be seen in Fig.~\ref{fig:funcDeriv}.

As shown in Fig.~\ref{fig:densVelsSigmas}, in contrast to $\Delta v_r$, $\Delta \sigma_{v_{\p}}$ is a roughly constant function of $r$ regardless of void size or model of gravity. In theory then one might expect this to induce changes in $\xi_{2}$ at all radii. Nevertheless, as shown in Fig.~\ref{fig:funcDeriv}, the effects of changes to the velocity dispersion are principally focused around the void edge. To understand why the signal is suppressed at larger radii, we note that in this region $\delta^{r}(r)$, $v_{r}(r)$, $\sigma_{v_{\p}}(r)$ are all effectively constant. In this limit, (\ref{eq:sigmaFD}) can be simplified to read (using subscript 0 to denote variable constant values) 
\bea
    \Delta_{\sigma_{v_{\p}}} \delta^{s} \approx \frac{\mathcal{H}(1+\delta^{r}_{0})(\Delta \sigma_{v_{\p},0})}{\sqrt{2 \pi} \sigma_{v_{\p},0}^2} \exp\left(-\frac{v_\p^2}{2 \sigma_{v_{\p},0}^2}\right)v_\p\  \bigg\vert^{+\infty}_{-\infty}= 0.\hspace{0.5cm}
\eea
We therefore expect that $\Delta_{\sigma_{v_{\p}}} \xi_{2}(s)$ will be very close to zero except where at least one of $\delta^{r}(r)$, $v_{r}(r)$, $\sigma_{v_{\p}}(r)$, or $\Delta \sigma_{v_{\p}}$ is changing rapidly with radial position in the vicinity of the associated $r \sim s$. The functions with the biggest potential impact in this regard are $\Delta \sigma_{v_{\p}}$ and $\delta^{r}$, as they both enter (\ref{eq:sigmaFD}) as overall multiplicative factors. In Fig.~\ref{fig:densVelsSigmas}, we find $\Delta \sigma_{v_{\p}}$ in both theories of gravity and both void populations is almost constant, while $\delta^{r}$ changes rapidly near the void edge. This in effect tethers the changes induced in $\xi_{2}$, by $\Delta \sigma_{v_{\p}}$ to this region, despite the fact that $\Delta \sigma_{v_{\p}}$ remains non-zero far outside of the void.

Examining this void edge region, increasing $\sigma_{v_{\p}}$ increases the amount of ``shuffling" of matter between neighboring radial bins which occurs during the move from real to redshift space. Near the void edge however, this shuffling is highly asymmetrical between the void interior and exterior. The void interior is mostly empty, and thus has very little matter which can stream out. The void exterior by contrast, has $(1+\delta^{r}) \sim 1$, and thus has some of its own matter shifted into the void while receiving almost none in return from the void center in the move from real to redshift space. This phenomena is again most pronounced along the line of sight, and increasing $\sigma_{v_{\p}}$ increases the severity. This means that $\Delta_{\sigma_{v_{\p}}} \delta^{s}$ will be positive for $s$ values within the void interior, negative for those $s$ values just outside of the void edge, and 0 for large $s$ values, with $\Delta{\sigma_{v_{\p}}} \xi_{2}$ following the same phenomenology.

$\Delta \delta^{r}$  is small compared to $(1+\delta^{r}\vert_{GR})$, as shown in Fig.~\ref{fig:densVelsSigmas}. As such, $\Delta \delta^{r}$ plays only a minor role in determining the total $\Delta \xi_{2}$ through (\ref{eq:delFD}). 

The discussion above outlines why increases in the radial velocity and velocity dispersion due to the modifications to gravity combine to give only minimal or no effect on $\xi_2$ in most instances. In summary, $\Delta v_{r}$ and $\Delta \sigma_{v_{\p}}$ have opposing effects on $\xi_{2}$. Increases to $v_r$ drive material outwards while increases to $\sigma_{v_{\p}}$ lead to the net movement of material inwards from the overdensity at the void edge. The end result is cancellation between $\Delta_{\sigma_{v_{\p}}} \xi_{2}$ and $\Delta_{v_{r}} \xi_{2}$, leading to the total $\Delta \xi_{2}$ being much smaller than the individual changes to $v_{r}$ and $\sigma_{v_{\p}}$. An exception to this  occurs in the large N1 voids at intermediate scales, $\sim 35-65 \mathrm{Mpc/h}$. In N1, $\Delta v_{r}$ is long ranged, which similarly extends the spatial range of $\Delta_{v_{r}} \xi_{2}$ and lessens its cancellation against $\Delta_{\sigma_{v_{\p}}} \xi_{2}$ - yielding the observed $\Delta \xi_{2}/ \Delta \xi_{2,GR} \sim 8\%$ over this range. In both F5 void populations, there is no significant difference from GR. The effects of the MG model are constrained to the void edge region where there is a direct cancellation, discussed above.

We have focused on the behavior of the quadrupole moment $\xi_{2}$ 
justified by noting that it is the first non-zero multipole moment induced entirely by redshift space distortions, as $\xi_{0} = \delta^{r}$, $\xi_{\ell\geq1}\equiv 0$ for stacked real space voids. We note that the functional derivative approach can be applied similarly to higher order even moments by the appropriate substitution of the Legendre polynomials in (\ref{eq:generalFD}) to isolate the effect of $\Delta \delta^{r}$, $\Delta v_{r}$, and $\Delta \sigma_{v_{\p}}$ on $\Delta \xi_{0}(s)$, $\xi_{4}(s)$, or any higher order even multipole moment (all odd moments are still identically 0 by symmetry). We find that the impact of the modified gravity model on the hexadecapole, $\xi_4$,  is much smaller than that for the quadrupole largely due the greater spatial oscillation in the higher order Legendre polynomial. As such, $\xi_{2}$ provides the best case scenario to test modified gravity with void multipole moments. 

\subsection{Fitting $\xi_2$ with the $\bold{(\beta, \ \sigma_{0})}$ Model}
\label{sec:betaSigma}

\begin{table}[t!]
\begin{tabular}{|C{3.5em}|C{7.em}|C{7.em}|C{7.em}|}

\cline{2-4}
\multicolumn{1}{c|}{}& \multicolumn{2}{c|}{Fit from $\xi_2$}&{Best fit from $v_r$}
\\
\hline
Model    &  $\sigma_0$  (km/s) & $\beta$ & $\beta$   
\\ \hline
GR & $278 \pm 17$  & $0.36 \pm 0.03 $        & $0.36$                 
\\ \hline
F5 & $287 \pm 18$ &$0.37 \pm 0.03$          & $0.37$                 
\\ \hline
N1 & $304 \pm 18$ &$0.39 \pm 0.03$          & $0.39$             
\\ \hline
\end{tabular}

\caption{
Comparison of the constraints on the $(\beta,\sigma_{0})$ model fit to the void quadrupole $\xi_{2}(s)$ in each theory of gravity, and a comparison with the best fit $\beta$ directly obtained from the void velocity profile, $v_{r}$.}
\label{tab:betaTable}
\end{table}

In this section we consider the application of the measured void quadrupole to constrain the underlying cosmological theory. Specifically, we consider the approach proposed in \cite{2019MNRAS.483.3472N} to use the quadrupole to constrain the cosmological parameter $\beta = f/b$, where $f=d\ln \delta_{DM}/d\ln a$ is the dark matter logarithmic growth rate and $b$ is the galaxy tracer bias. 

Within the context of GR and linear theory, the void radial velocity profile is given by
\bea
   v_r(r) = -\frac{\beta}{3}\mathcal{H} \Delta(r)r
   \label{eq:vrlin}
\eea
where $\Delta(r)$ is the average density within a radius $r$,
\bea
\Delta(r) \equiv \frac{1}{\frac{4}{3} \pi r^3}\int_0^r 4 \pi \delta^{r}(r') {r'}^2 dr'.
   \label{eq:linThy}
\eea
Here $\beta=f/b$ is employed instead of $f$ since we calculate $\Delta(r)$ using the galaxy number density contrast $\delta^{r}$ instead of the underlying dark matter density contrast itself. 

Using the simulations, in which we have complete knowledge of all tracer velocities, we are able to explore the connection between $\beta$ and $v_r$  directly. We find that for the above median size voids the linear relationship in (\ref{eq:vrlin}) holds well, however it is not consistently followed in the smaller void sample. For this reason we focus this analysis on the larger void sample. In Table~\ref{tab:betaTable}, we show the best fit values of $\beta$ inferred from a fit of the average radial velocity, $v_{r}$, using (\ref{eq:vrlin}) for the above-median size void population for each theory of gravity. Modifications to gravity increase $v_{r}$, which in the  context of  (\ref{eq:vrlin}), lead to an increase in the inferred $\beta$. 

While the simulations give us direct access to the velocity information, real world  observational programs do not. However, we can still use the measured quadrupole moment $\xi_{2}$ along with the GSM to constrain the cosmological parameter $\beta$. We use a modified version of the GSM where only an average void density profile $\delta^{r}$ is assumed along with two cosmological parameters, $\beta$ and $\sigma_{0}$, which replace direct knowledge of both functions containing velocity information. An effective $v_{r}$ is then constructed using $\beta$ in conjunction with (\ref{eq:linThy}), while the velocity dispersion is set to a constant effective value $\sigma_{v_{\p}}(r) = \sigma_{0}$. This $(\beta,\sigma_{0})$ model is a simple parameterization used to capture the dynamics within void environments without having to have an exact knowledge of the two functions containing velocity information.

To assess the utility of this approach, we consider the mean $\xi_2$ measured across the 100 simulations, and estimate the signal covariance from the uncertainties from a Jackknife from a single realization with volume of $\sim (1\mathrm{Gpc/h})^{3}$, comparable to the volume expected to be sampled by DESI at $z=0.5$ \cite{Font_Ribera_2014}. 

\begin{figure}[!t]
\includegraphics[width=1.0\linewidth]{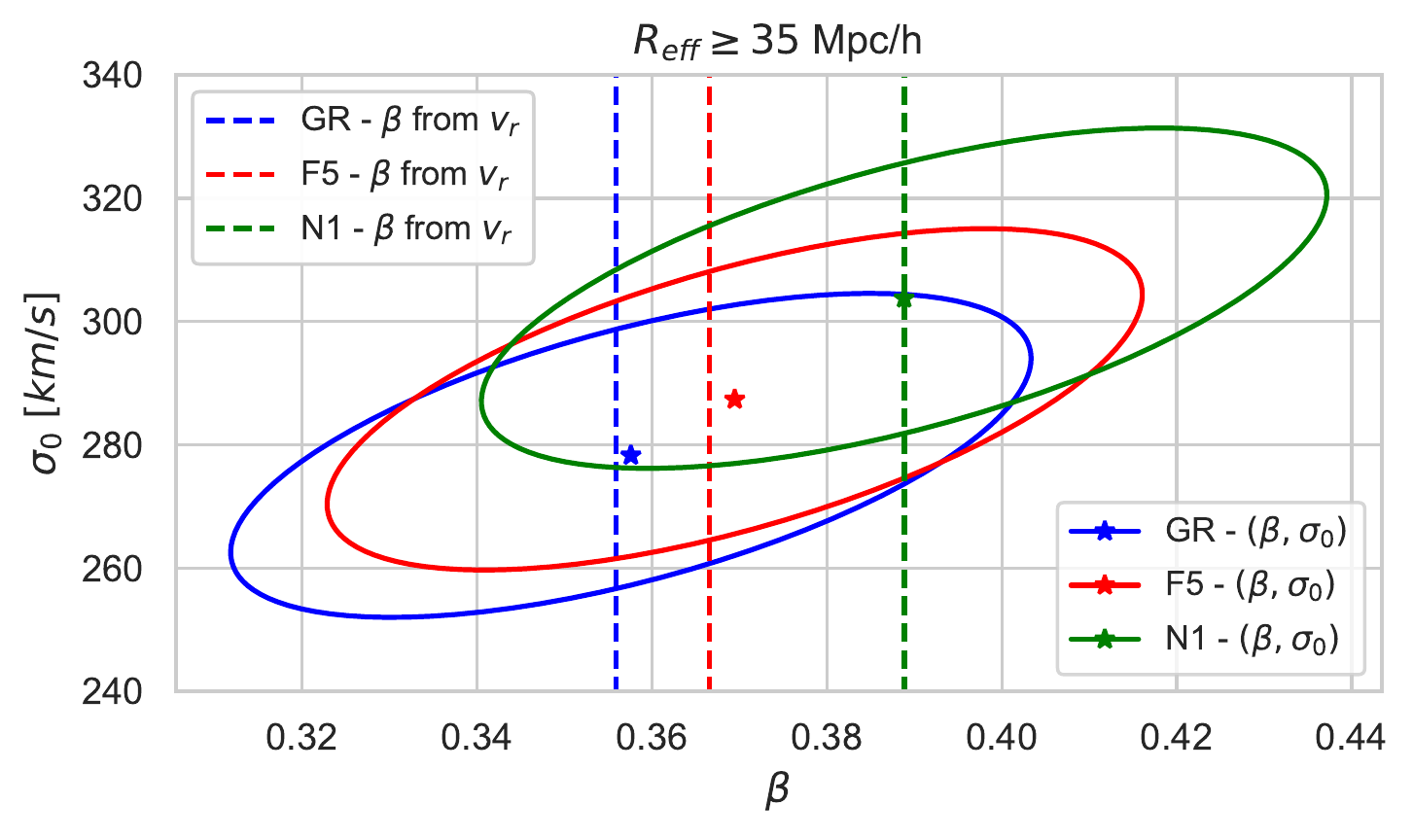}
\caption{The linear growth rate dependent parameter, $\beta$, calculated from $v_{r}$ directly using linear theory [dashed lines], compared with the best fit [star] and the 68\% confidence ellipses [full line] from $\xi_2$ modeled using the GSM with the ($\beta,\sigma_{0})$ parameterization, in GR [blue], F5 [red], and N1 [green]. The model is fit to the mean quadrupole from 100 realizations for the above-median size void sample with confidence ellipses derived from statistical uncertainties estimated for 1 realization, with volume $(1.024 \mathrm{Gpc/h})^3$. }
\label{fig:confidenceEllipses}
\end{figure}

Figure~\ref{fig:confidenceEllipses} shows the best fit values and the corresponding 68\% confidence interval for $\beta$ and $\sigma_{0}$ recovered from the joint 2D fit to $\xi_{2}$. The plot also shows the value of $\beta$ recovered directly in each theory from fitting the average void $v_r$ profile using (\ref{eq:vrlin}). The 1D projected constraints on the $(\beta,\sigma_0)$ fit parameters for the three gravity models are summarized in Table~\ref{tab:betaTable}

The 1D value of $\beta$ recovered by each method match, showing that the $\xi_2$ fit, including $\sigma_0$, does not bias the recovered value of $\beta$. The  values of $\sigma_{0}$ do not match the asymptotic values of $\sigma_{v_{\p}}$ shown in Fig.~\ref{fig:densVelsSigmas}. This is due to the fact that, as previously discussed, as long as $\sigma_{v_{\p}}$ is constant, $\xi_{2}$ is fairly insensitive to its exact value away from the void center. Thus, the value $\sigma_{0}$ takes is more sensitive to dynamics near the void edge, and less to the actual asymptotic value of $\sigma_{v_{\p}}$. It's to be expected therefore that the value of $\sigma_0$ should be lower than the asymptotic value far from the void center. This is consistent with what we find. For GR, for example, $\sigma_{v_\p}$ asymptotes to a value of $305$ km/s while $\sigma_0$ is $278$ km/s, while for F5 the values are $318$ km/s and $287$ km/s respectively.

While N1 and F5 have larger predicted values of $\beta$ than GR, the values obtained for the three models are indistinguishable within the estimated uncertainties at the 68\% confidence level considering the volume observable within this $z=0.5$ redshift slice. This opens up the possibility, however, that combining data from multiple redshift slices (to increase the observational volume and lower the level of statistical uncertainties) could provide a way to increase sensitivity and allow the application of the void quadrupole to distinguish between models in this way.

\section{Conclusions}
\label{sec:conc}

In this paper we utilize the large scale, high precision N-body MG-GLAM simulations \citep{Klypin:2018MNRAS.478.4602K.GLAM,Ruan_2022,Hernandez-Aguayo:2021arXiv211000566H.MGGLAM.DGP} of GR, $f(R)$, and nDGP gravity to compare real space dynamical properties of voids and the resulting redshift space quadrupole moments, as might be measured from upcoming spectroscopic galaxy survey such as from the DESI, Euclid and Roman experiments. This work builds on of our previous work  \cite{PhysRevD.104.023512}, where scale dependent effects were observed in void peculiar velocities within $f(R)$ gravity.

While the properties of voids can be investigated using dark matter halos as biased tracers of the underlying dark matter distribution itself, in order to compare with observations, we consider how the void statistics are modified by the inclusion of a HOD. We find that it is vital to include the effects of an HOD when identifying voids and calculating the resulting statistics. The application of a HOD significantly effects the void size function in the $f(R)$ model. The $f(R)$ model predicts more halos for given initial conditions, however the HOD equalizes the void size function so that it becomes consistent with GR. The HOD approximately normalizes the total number of tracers between the two theories when assigning mock galaxies, which in turn equalizes the number of voids identified with a watershed void finder.

The radial velocity, $v_{r}$, in all models has the same qualitative form, rising to a peak around the void edge and then tending to zero as one moves out away from the void. In both nDGP and $f(R)$ theories the radial velocity is enhanced relative to GR.  In $f(R)$ the enhancement is most pronounced in smaller voids. While the magnitude of the increase in $v_{r,peak}$ for $f(R)$ in HOD identified voids is suppressed relative to those identified from all halos, the difference between GR and the two MG models remains present in voids of all sizes. The  relative reduction in the velocity for $f(R)$ results from the HOD preferentially populating larger mass halos with mock galaxies, where these larger halos experience more chameleon screening compared to their smaller counterparts. In the N1 nDGP gravity model there is less scale dependence in the enhancements to the void peculiar velocity, $v_{r}$, in accordance with the linearized nDGP field (\ref{eq:linVarphi}). The addition of a HOD has less impact; while the HOD still preferentially populates more massive halos, the shift symmetry present in (\ref{eq:fieldVarphi}) allows halos of all sizes to respond in a similar fashion to slowly varying background fields which source the fifth force (unlike $f(R)$ gravity). Voids identified with halos and after the HOD is applied therefore yield similar enhancements to $v_{r}$ from N1 gravity.

The velocity dispersion, $\sigma_{v_{\p}}$, also has a common qualitative form in all models, being largely constant as one moves out from the void edge. The velocity dispersion is enhanced relative to GR in both the $f(R)$ and nDGP models, with $\sim$5\% and 10\% enhancements relative to GR respectively, and in both cases has little dependence on void size.

Given the scale dependent enhancements observed in $v_{r}$ in the $f(R)$ model, we consider two size-based samples in our analyses, those above and below the median radius of $R_{\mathrm{eff}}=35 \mathrm{Mpc/h}$.

We perform a detailed analysis of the void quadruple moment, $\xi_{2}$, to understand how sensitive redshift space statistics are to changes in the void dynamical properties within modified gravity. While we find increases in void velocity statistics of $\sim5\%-10\%$ relative to GR in both void size populations and both modified theories, we find that changes to $\xi_{2}$ in both F5 or N1 are far more muted. For F5, the differences in velocities do not yield comparable differences in $\xi_{2}$ to that in GR. In N1, we find a difference for the large void population of approximately $8\%$ in $\xi_{2}$ in a limited region, 35-65Mpc/h away from the void center.

We utilize the Gaussian streaming model (GSM) to determine the origins of these findings. We first study the accuracy of the GSM in reproducing the redshift space void quadruple moment for modified gravity models. We find that the model is highly accurate in reproducing $\xi_{2}$, in both large and small-size void samples, and for both theories of modified gravity. We apply the model to understand how differences in the velocities and densities propagate to cause differences in $\xi_{2}$ in both modified gravity theories. Specifically, we isolate the changes induced in $\xi_{2}$ individually by $\delta^{r}$, $v_{r}$, and $\sigma_{v_{\p}}$ using the first order functional derivatives of the GSM.

In the functional derivative approach, the effects of $\Delta v_r$ and $\Delta \sigma_{v_\p}$ on $\xi_{2}$ are opposite in sign and therefore act to counter each other in the same spatial region. Using this approach further, we find that changes to $\xi_{2}$ induced by changes to the line of sight velocity dispersion $\Delta \sigma_{v_{\p}}$ are found to be limited to the void edge region, despite that fact that the velocity dispersion continues to be enhanced at distances much further from the void center. The basis for this is that changes to $\xi_2$ cannot be driven by a constant $\Delta \sigma_{v_{\p}}$ unless the other GSM functions (primarily $\delta^{r}$) simultaneously have non-zero radial gradients. Far from the void, all three GSM functions are slowly varying, which restricts changes induced in $\xi_2$ by the increased velocity dispersion to the void edge, where $\delta^{r}$ does have a large radial gradient. By contrast, the changes to $\xi_{2}$ induced by changes to the radial velocity $\Delta v_{r}$ are not, in principle, confined to the area surrounding the void edge but instead extend throughout the entire region where $\Delta v_{r}$ is non-zero. The spatial extent of $\Delta v_{r}$ in both our $f(R)$ and nDGP model can be qualitatively related to the amount of screening present within the linearized field equations, (\ref{eq:linfR}) and (\ref{eq:linVarphi}) respectively. For $f(R)$, (\ref{eq:linfR}) experiences Yukawa screening, and thus, $\Delta v_{r}$ is short-ranged and fails to extend far beyond the void edge. This in turn means that the changes induced in the quadrupole by $\Delta v_{r}$ and $\Delta \sigma_{v_{\p}}$, are {\it both} confined to the same spatial region near the void edge, and being of opposite sign, cancel heavily and leave $\xi_{2,F5}$ largely unchanged from $\xi_{2,GR}$. On the other hand, in nDGP,  (\ref{eq:linVarphi})  is unscreened at the linear level. This means that in nDGP gravity, particularly in the above median size voids, that $\Delta v_{r}$ has a  large spatial range. For nDGP therefore, the changes to $\Delta \xi_{2}$ induced by $\Delta v_{r}$ and $\Delta \sigma_{v_{\p}}$  are {\it not} confined to the same spatial region, and avoid substantial cancellation. This leads to differences arising between $\xi_{2,N1}$ and $\xi_{2,GR}$ in the above median size void population for $s\sim35-65\mathrm{Mpc/h}$.

We translate differences between $\xi_{2}$ in each gravitational theory into constraints on the cosmological growth rate parameter $\beta$ through the use of linear theory and the $(\beta, \sigma_{0})$ model. We find that within the large void population, $v_{r}$ can be accurately fit with linear theory, resulting in increased values of $\beta$ for each modified gravity theory over the GR value. Although $v_{r}$ is not an observable statistic, we can use it in the simulations to verify that a two-parameter fit to $\xi_{2}$ within the context of the GSM, with $v_{r} \propto \beta$ and $\sigma_{v_{\p}}(r) = \sigma_{0}$, can recover, in an unbiased manner, the same value of $\beta$ as that recovered from $v_{r}$ directly. Increased values of $\beta$ are recovered in both $f(R)$ and nDGP gravitational scenarios.

When statistical uncertainties are computed corresponding to an observational volume of $\sim 1 (Gpc/h)^3$ at $z=0.5$, we find the values of $\beta$ recovered from the $\xi_{2}$ fit in each modified gravity theory lie within the projected 1D error bars of the GR value. Applying this approach to an analysis of multiple redshift slices, akin to the full redshift range probed by upcoming spectroscopic surveys, will no doubt provide greater distinguishing power, although this is beyond the scope of this paper.

Our results show how the theoretical GSM model works remarkably well at modeling void dynamics in theories beyond GR. It also details, however, the challenges  present in utilizing redshift space distortion data around cosmic voids to constrain the properties of gravity. The work also demonstrates the importance of considering the statistics derived from realistic HOD-derived tracers rather than the dark matter halos directly. We anticipate that the results from this paper will have broader applicability for accurately determining the constraining potential of cosmic voids for other non-standard cosmological models.

\section*{Acknowledgements}

We wish to thank Baojiu Li and Cheng-Zong Ruan for kindly providing the MG-GLAM simulation products, and providing advice and assistance in their use. The work of Christopher Wilson and Rachel Bean is supported by NSF grant AST 2206088, NASA ATP grant 80NSSC18K0695, and NASA ROSES grant 12-EUCLID12-0004.

\appendix
\label{sec:app}

\section{Rescaled analysis}
\label{app:rescaled}

In the main analysis we present results where the radial coordinate of the stacked voids is given in terms of the standard comoving distance, with units of comoving Mpc/h.  Due to the symmetry properties of voids, there is a commonality in their density profiles when they are stacked using an alternative radial coordinate rescaled by each void's effective radius \cite{Hamaus_2014}. This rescaling allows voids to be conveniently stacked with each other in a way that draws on the void similarities, and can alleviate effects caused by differences in void size when stacking voids of largely different $R_{\mathrm{eff}}$. 

In this appendix, we present results using this rescaled coordinate system as the partner results to those in the main text. The rescaled coordinates are defined as $\tilde{r}=r/R_{\mathrm{eff}}$ for physical space and  $\tilde{s}$ in redshift space variables, with the relevant transformation being,
\begin{equation}
\vec{\tilde{s}}=\vec{\tilde{r}}+\frac{\vec{v} \cdot \hat{l}}{\mathcal{H} R_{\mathrm{eff}}}.
\end{equation}
When we perform the analysis binning tracers using the rescaled coordinates our definition of $\delta^{r}$ and $\delta^{s}$ slightly changes to reflect that for voids of differing $R_{\mathrm{eff}}$, spherical slices at equal $\tilde{r}$ or $\tilde{s}$ are different sizes. Interpreting the term $N_v V(r)$ in the denominator of (\ref{eq:deltaRUnscaled}) as the ``total volume averaged over", we use 
\begin{equation}
\delta^{\tilde{r}} (\tilde{r}) =\frac{N_{vh}(\tilde{r})}{\bar{n}_{h} \sum_{voids} V_{i}(\tilde{r}) } -1,
\label{eq:deltaRRescaled}
\end{equation}
and
\begin{equation}
\delta^{\tilde{s}} (\tilde{s},\mu_{s}) =\frac{N_{vh}(\tilde{s},\mu_{s})}{\bar{n}_{h} \sum_{voids} V_{i}(\tilde{s},\mu_{s}) } -1,
\label{eq:deltaSRescaled}
\end{equation}
where $V_{i}(\tilde{s},\mu_{s})$ is the volume (in unscaled space) of the bin at $(\tilde{s},\mu_{s})$ in the $i^{th}$ void. For rescaled voids, we use 30 radial bins extending from $0$ to $3R_{\mathrm{eff}}$ away from the void center, with the same angular binning as used in the unscaled analysis.

The Gaussian streaming model is also easily adapted for the purpose of analyzing redshift space distortions around rescaled voids. The only real differences from (\ref{eq:randCoord}) are that $v_{r} \rightarrow v_{r}/R_{\mathrm{eff}} \equiv \tilde{v}_{r}$, $v_{\p} \rightarrow v_{\p}/R_{\mathrm{eff}} \equiv \tilde{v}_{\p}$, which then requires $\mathcal{P} \rightarrow \tilde{\mathcal{P}}$ with $\sigma_{v_{\p}} \rightarrow \sigma_{v_{\p}}/R_{\mathrm{eff}}\equiv \tilde{\sigma}_{v_{\p}}$. It should be noted that in computing these quantities, each void re-scales the velocities of its own tracers by its own $R_{\mathrm{eff}}$ before averaging to calculate $\tilde{v}_{r}$ and $\tilde{\sigma}_{v_{\p}}$

\begin{figure*}[t!]
\includegraphics[width=1.0\linewidth]{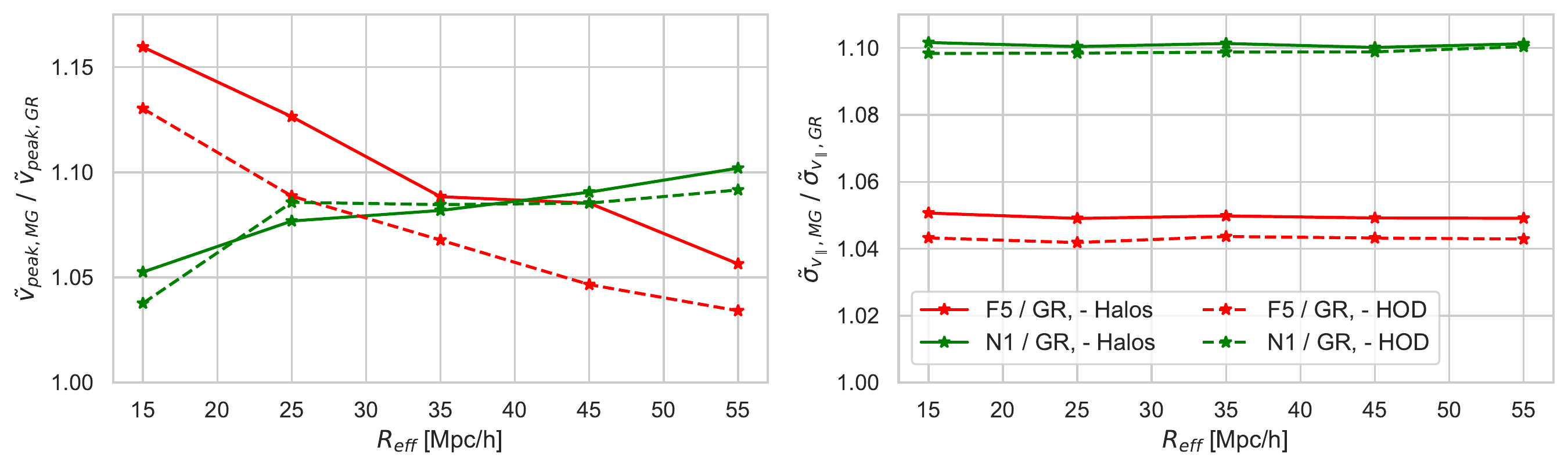}
\caption{Left: Ratio of the peak rescaled radial velocity in F5 [red] and N1 [green] to that in GR as a function of void effective radius, $R_{\mathrm{eff}}$, for voids identified from halos [solid] and from halos containing mock galaxies identified by the HOD [dashed]. Right: Ratio of the average asymptotic value of the rescaled velocity dispersion, at $\tilde{r} \gtrsim 1.5$, as a function of $R_{\mathrm{eff}}$.}

\label{fig:ratiosRescaled}
\end{figure*}

\begin{figure*}[t!]
\includegraphics[width=1.0\linewidth]{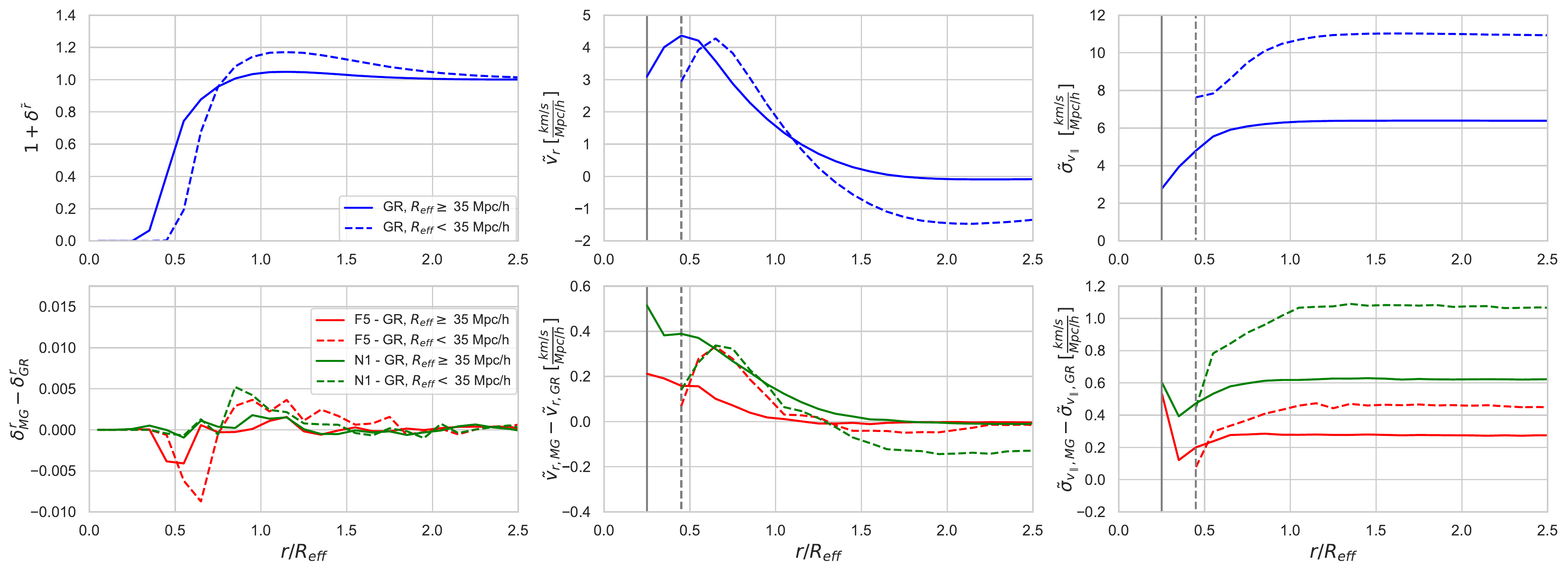}
\caption{Top Row: Void density [left], rescaled radial velocity [center] and rescaled velocity dispersion [right] profiles in GR for voids identified in halos with HOD-identified tracers. The void population is split into those with void size above [solid] and below [dashed] $R_{\mathrm{eff}}$=35Mpc/h, the median void size found in each of the three theories. Gray lines indicate the first radial bin below which no tracers are found. Bottom Row: Differences between F5 [red] and N1 [green] quantities with respect to those in GR.}
\label{fig:densVelsSigmasRescaled}
\end{figure*}

\begin{figure*}[!t]
\includegraphics[width=1.0\linewidth]{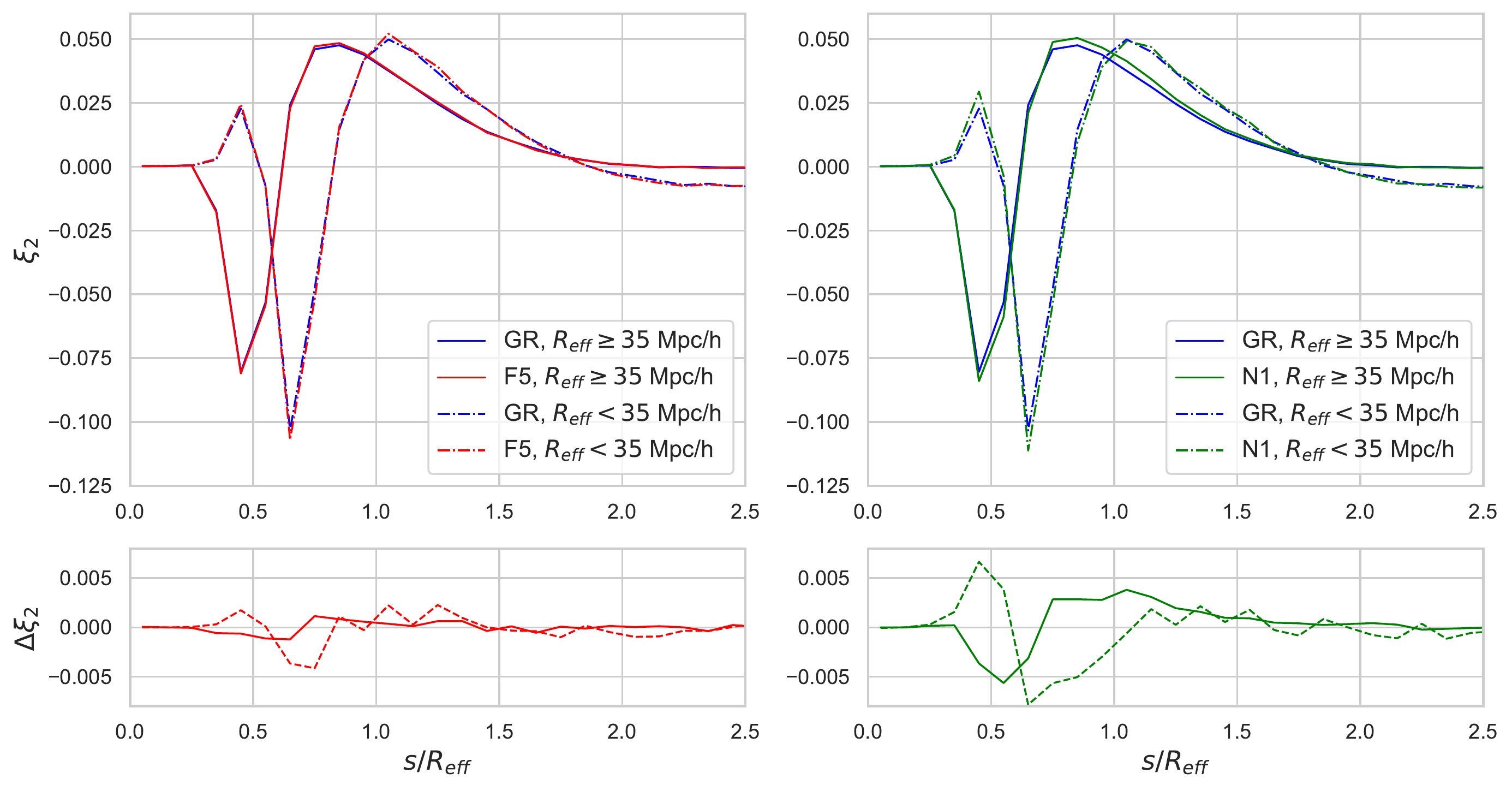}
\caption{Top Row: The void quadrupole moment $\xi_{2}$  at $z=0.5$ for GR [blue] compared against F5 [left, red] and N1 [right, green] in rescaled voids both below [dashed] and above [solid] the median void size. Bottom Row: Difference in $\xi_{2}$ between GR and each modified gravity theory in small [dashed] and large [solid] voids.}
\label{fig:quadDataRescaled}
\end{figure*}

\begin{figure*}[!t]
\includegraphics[width=1.0\linewidth]{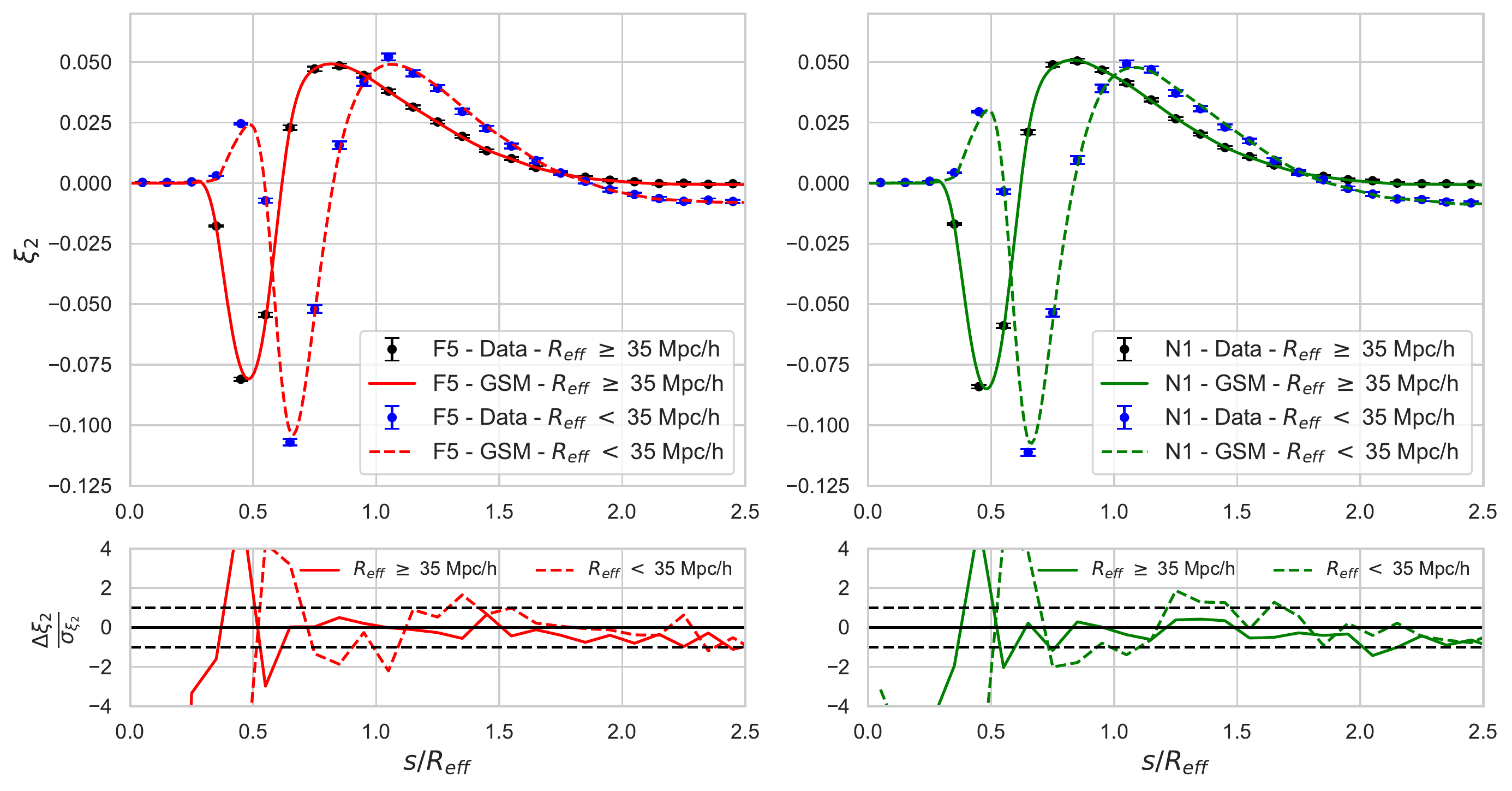}
\caption{Top: Theoretical value of $\xi_{2}$ in rescaled voids both above [dashed] and below [solid] the median void size in F5 [left, red] and N1 [right, green] calculated using the Gaussian Streaming Model (GSM) compared to average $\xi_{2}$ from the simulations. Bottom: Difference between the GSM-derived quadrupole and that from the simulated data, $\Delta\xi_2=\xi_{2,GSM}-\xi_{2,Data}$, with respect to the statistical uncertainties for the model, $\sigma_{\xi_{2}}$.}
\label{fig:quadThyRescaled}
\end{figure*}

\begin{figure*}[!t]
\includegraphics[width=1.0\linewidth]{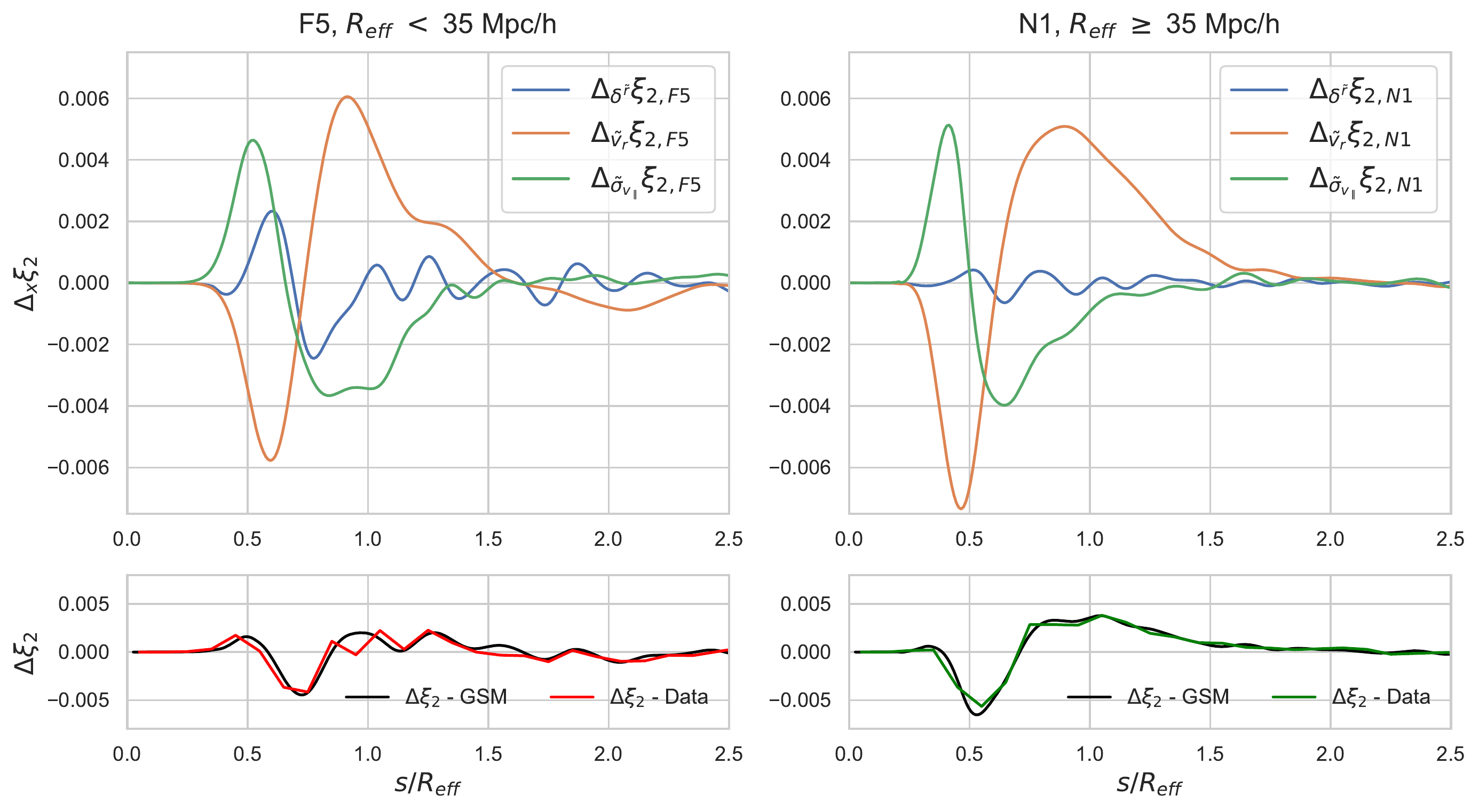}
\caption{Two scenarios are considered: [Left] Below median size rescaled voids in F5 and [Right] above median size rescaled voids in N1. Top Row: The changes to $\xi_{2}$ relative to GR induced independently by $\Delta \delta^{\tilde{r}}$ [blue], $\Delta \tilde{v}_{r}$ [orange], and $\Delta \tilde{\sigma}_{v_{\parallel}}$ [green] as calculated using GSM functional derivatives. Bottom Row: Comparison of the differences,  $\Delta\xi_{2}$, between the modified gravity model and GR obtained from the simulated data in F5 [left,red] and N1 [right, green] and the summed GSM integrated functional derivatives [black].}
\label{fig:funcDerivRescaled}
\end{figure*} 

\begin{figure}[!t]
\includegraphics[width=1.0\linewidth]{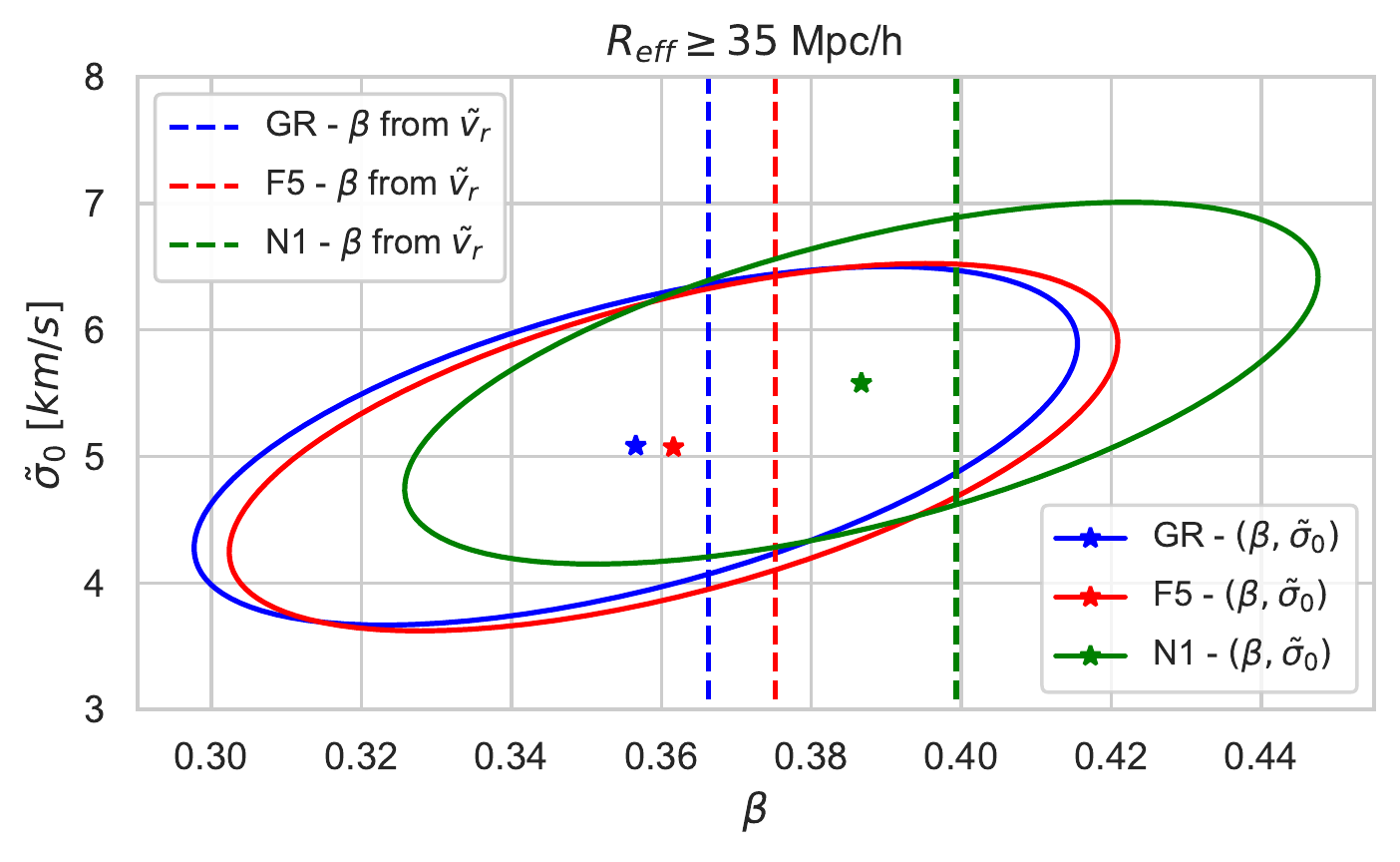}
\caption{The linear growth rate dependent parameter, $\beta$, calculated from $\tilde{v}_{r}$ directly using linear theory [dashed lines], compared with the best fit [star] and the 68\% confidence ellipses [full line] from $\xi_2$ modeled using the GSM with the ($\beta,\tilde{\sigma}_{0})$ parameterization, in GR [blue], F5 [red], and N1 [green]. The model is fit to the mean quadrupole from 100 realizations for the above-median size void sample with confidence ellipses derived from statistical uncertainties estimated for 1 realization, with volume $(1.024 \mathrm{Gpc/h})^3$. }
\label{fig:confidenceEllipsesRescaled}
\end{figure}

Figures~\ref{fig:ratiosRescaled}-\ref{fig:confidenceEllipsesRescaled} show the rescaled counterparts to Figs~\ref{fig:ratios}-\ref{fig:confidenceEllipses}. The results obtained using the rescaled quantities are all consistent with those presented in the main text using unscaled quantities, although there are some differences caused by the rescalings which we comment on here, below.

Figure \ref{fig:ratiosRescaled} shows the same patterns as shown in Fig.~\ref{fig:ratios}. Comparing the two figures, we can see that the effects caused by the transition from voids selected from halos to instead HOD mock galaxies are more clear when using rescaled variables.

Figure~\ref{fig:densVelsSigmasRescaled} shows the the rescaled real space density profile, $\delta^{\tilde{r}}(\tilde{r}$,  the rescaled void radial velocity $\tilde{v}_{r}$, and the rescaled velocity dispersion profile. In Fig.~ \ref{fig:densVelsSigmas}, we saw that the unscaled $v_{r}$ displayed a strong dependence on void size in $f(R)$, with its peak value increasing almost linearly with $R_{\mathrm{eff}}$ in accordance with linear theory. Concurrently,  $\sigma_{v_{\p}}$ was independent of $R_{\mathrm{eff}}$ and plateaued to the same value for both small and large voids. In Fig.~ \ref{fig:densVelsSigmasRescaled}, with rescaled quantities, these properties are reversed. Dividing by a factor of $R_{\mathrm{eff}}$ effectively makes $\tilde{v}_{r}$ scale independent, with differences in $\tilde{v}_{r}$ instead coming from differences in $\delta^{\tilde{r}}(\tilde{r})$ and not differences in $R_{\mathrm{eff}}$ between the two void populations. Whereas previously $\sigma_{v_{\p}}$ was scale independent, dividing by$ R_{\mathrm{eff}}$ introduces scale dependence in $\tilde{\sigma}_{v_{\p}}$, causing smaller voids, which divide by a smaller $R_{\mathrm{eff}}$, to have larger values of $\tilde{\sigma}_{v_{\p}}$ than larger voids. 

The bottom row of Fig.~\ref{fig:densVelsSigmasRescaled} shows the spatial extent of $\Delta \tilde{v}_{r}$ is consistent with that of the unscaled $\Delta v_{r}$. $\Delta \tilde{v}_{r}$ extends out past $1.5 \tilde{r}$ in the large N1 voids, and is more confined, to $\tilde{r}<1$, in both the large and small radius F5 void samples.

Figure \ref{fig:quadDataRescaled} shows that the trends in the void quadrupole moments for rescaled coordinates are the same as seen for unscaled coordinates in Fig.~ \ref{fig:quadData} in both the large and small void populations in each of the three theories of gravity considered.

Figures~\ref{fig:quadThyRescaled} and \ref{fig:funcDerivRescaled} demonstrate that  
the GSM approach and functional derivative analysis can be accurately used to predict the quadrupole using rescaled quantities across both void populations and all theories of gravity (GR not shown).

In order to reproduce Figure~\ref{fig:confidenceEllipses} for rescaled voids, we fit $\tilde{v}_{r}$ with linear theory. The re-scaling slightly modifies (\ref{eq:linThy}), so that for $\tilde{v}_{r}$, we  have
\bea
   \tilde{v}_r(\tilde{r} &=& -\frac{\beta}{3}\mathcal{H} \Delta_{gal}(\tilde{r}) \tilde{r} \\
   \Delta_{gal}(\tilde{r}) &\equiv& \frac{1}{\frac{4}{3} \pi \tilde{r}^3}\int_0^{\tilde{r}} 4 \pi \delta_{gal}^{r}(\tilde{r}') \tilde{r}'^2 d\tilde{r}'
   \label{eq:linThyTilde}
\eea
where $\delta^{\tilde{r}}(\tilde{r})$ is computed as in (\ref{eq:deltaRRescaled}).

Figure \ref{fig:confidenceEllipsesRescaled} shows the confidence ellipses for $(\beta,\tilde{\sigma}_{0})$ from rescaled voids in GR, F5, and N1, as well as the best fit values of $\beta$ calculated directly from fitting  (\ref{eq:linThyTilde}) to the true rescaled radial velocity profile $\tilde{v}_{r}$. There is only a slight difference between the vertical lines shown in Fig.~\ref{fig:confidenceEllipsesRescaled} and those in Fig.~ \ref{fig:confidenceEllipses}, indicating consistency in the methods used to calculate $v_{r}$ and $\tilde{v}_{r}$ from linear theory. When comparing scaled versus unscaled, the recovered $\beta$ from the $(\beta,\sigma_{0})$ fit is closer to the ``true" value in unscaled voids (Fig.~\ref{fig:confidenceEllipses}), although both are well within the 1-$\sigma$ confidence ellipse for all three theories of gravity. The difference arise due to the $\sigma_{v_{\p}}(r)=\sigma_{0}$ assumption providing a better fit to the true $\sigma_{v_{\p}}(r)$ profile in unscaled voids compared to rescaled ones (Fig. ~\ref{fig:densVelsSigmas} vs. Fig.~\ref{fig:densVelsSigmasRescaled}). 

\clearpage

\newpage

\bibliographystyle{apsrev}
\bibliography{references2}

\end{document}